\documentclass[preprint,12pt]{aastex}

\def\zs{z_s}
\def\kpc{{\rm\;kpc}}

\def\munit{10^9M_\odot\kpc^{-2}}

\def\galax{J0414+053}
\def\snclus{J1004+411}
\def\abclus{ACO~1689}

\def\hover#1{\setbox0\hbox to 0pt{\hss$\scriptscriptstyle
             \rightharpoonup$}%
             #1\kern0.4ex\raise 1.5ex\box0\kern-0.01ex}
\def\btheta{{\hover\theta}}
\def\bbeta{{\hover\beta}}

\def\hovelr#1{\setbox0\hbox to 0pt{\hss$\scriptscriptstyle
              \leftharpoonup\!\!\!\!\rightharpoonup$}%
              #1\raise 1.5ex\box0\kern0.1ex}
\def\M{{\hovelr M}}

\begin{document}

\title{Meso-structure in three strong-lensing systems}

\author{Prasenjit Saha}
\affil{Institute for Theoretical Physics, University of Z\"urich, \\
       Winterthurerstrasse 190, 8057 Z\"urich, Switzerland}

\author{Liliya L.R. Williams}
\affil{Department of Astronomy, University of Minnesota, \\
       116 Church Street SE, Minneapolis, MN 55455}

\author{Ignacio Ferreras}
\affil{Physics Department, King's College London \\
       Strand, London WC2R 2LS, UK}

\begin{abstract}

We map substructure in three strong lensing systems having
particularly good image data: the galaxy lens MG~J0414+053 and the
clusters SDSS~J1004+411 and ACO~1689.  Our method is to first
reconstruct the lens as a pixelated mass map and then substract off
the symmetric part (in the galaxy case) or a projected NFW (for the
cluster lenses). In all three systems we find extended irregular
structures, or meso-structures, having of order 10\% of the total
mass.  In J0414+053, the meso-structure suggests a tidal tail
connecting the main lens with a nearby galaxy; however this
interpretation is tentative.  In the clusters the identification of
meso-structure is more secure, especially in ACO~1689 where two
independent sets of lensed images imply very similar meso-structure.
In all three cases the meso-structures are correlated with galaxies
but much more extended and massive than the stellar components of
single galaxies. Such extended structures cannot plausibly persist in
such high-density regions without being mixed; the crossing times are
too short. The meso-structures therefore appear to be merging or
otherwise dynamically evolving systems.

\end{abstract}

\keywords{gravitational lensing --- galaxies: clusters: individual
(SDSS J1004+411, ACO 1689)}

\section{Introduction}

It is a commonplace nowadays that astrophysical systems once thought
to be isolated and quiescent turn out to have complicated
substructures that are evolving now.  To give just two recent
examples: \cite{2006ApJ...642L.137B} discern two wraps of the
Sagittarius Dwarf galaxy as it merges into the halo of the Milky Way,
while in the Coma cluster \cite{2005A&A...443...17A} identify 17
different galaxy groups.

Substructure has long been of special interest in gravitational
lensing systems, because of the prospects of detecting dark matter
substructure. An early example is \cite{1995A&A...297....1B} who argue
that mass substructure in clusters is necessary to explain the
observed distribution and properties of arcs. Later \cite{1998MNRAS.295..587M}
noted that mass substructure (on much smaller mass and length scales that in the
case of clusters) readily explains the observed magnification
ratios in a galaxy lens, and more recently this topic has attracted
more attention.  There is general agreement that substructure and
microlensing are both present and both contribute to anomalous flux
ratios in galaxy lenses, but there is considerable debate as to whether
the expected CDM substructure can account for the observations
\citep{2002ApJ...572...25D,2002ApJ...567L...5M,2003MNRAS.345.1351E,
2004A&A...423..797B,2006MNRAS.367.1367A,2006MNRAS.368..599M}.
Such comparison of observed and model-predicted flux ratios is entirely
statistical; it does not attempt to map particular substructures, as the mass
of individual clumps is rather small, in the range $10^4-10^7 M_\odot$.
On the other extreme, some galaxies and clusters have substructures
comparable to the size of the whole lens.  For example, if a lens consists
of two or more distinct components of comparable mass, uncovering these
is relatively  straightforward
\citep{1998MNRAS.294..734A,1998AJ....116.1541A,2002ApJ...568..141G}.
The most elusive type of substructure is that which comprises roughly 10\%
of the lens mass, but has a considerable covering factor, or order of 50\%.
Until recently only tentative maps of such structures have been possible
\citep{2004AJ....128.2631W}.

In this paper we map and interpret substructure in three lensing
systems with exceptionally good lensing data.  Previous work has
identified substructure in all three systems, but here we will go
into more details.  The first system is the galaxy lens \galax,
which is a quad unique for the magnification information it provides.
\cite{2000ApJ...535..671T} resolved the lensed images using VLBI to
$\sim0.01''$, revealing multiple components, and hence supplying
tensor relative magnifications between the images. The second system
is cluster SDSS~\snclus, which has 14 images from four sources
\citep{2005ApJ...629L..73S} and even a time delay measurement
\citep{fohlmeister}.  The third system is the inner region of \abclus,
well known as an exceptionally rich lensing cluster, where
\cite{2005ApJ...621...53B} have now identified 30 separate lensed
sources giving rise to over 100 images in all.

\section{The galaxy lens MG \galax} \label{sys1}

\galax\ is a quad where each image has multiple components
$\sim0.01''$ apart.  The main lens is an early type galaxy G, but there
is also a second, fainter galaxy X that also contributes to the lens.
There are two independent sets of VLBI observations, with associated
models. \cite{2000A&A...362..845R} identify at least two distinct
components in the quadruply-imaged source, and fit the image positions
and flux ratios with two isothermal spheres plus external
shear. \cite{2000ApJ...535..671T} go even further, identifying four
source components and developing a lens model made up of basis
functions that fits the source substructure in detail.

Resolved image components provide information on magnification beyond
flux ratios.  Recall that the magnification tensor $\M$ relates a
source differential $d\bbeta$ and the corresponding image differential
$d\btheta$ by
\begin{equation}
d\btheta = \M d\bbeta .
\end{equation}
At another image from the same source there will be a different
magnification tensor $\M'$, leading to a different image differential
\begin{equation}
d\btheta' = \M' d\bbeta .
\end{equation}
Since the source differential is common we have
\begin{equation}
d\btheta' = \M' \M^{-1} d\btheta .
\label{relmag}
\end{equation}
For unresolved sources the flux ratio is $\det|\M'\M^{-1}|$. On the
other hand, resolved sources can measure the tensor relative
magnification $\M'\M^{-1}$.  This point was made by
\cite{1984ApJ...287..538G} and their Eq.~(2) is essentially our
Eq.~(\ref{relmag}).  They and later authors used the idea to
incorporate tensor magnification in models of B0957+561
\citep{1994MNRAS.270..457G,1996ApJ...464...92G} and B1422+231
\citep{2003AJ....126...29R}.

To the \cite{1984ApJ...287..538G} result we may add that explicit
calculation of the tensor magnification is not necessary.  If three
non-colinear features in the source are resolved in the images, they
provide two linearly independent source displacements, and hence the
relative tensor magnification is automatically constrained.  In
practice there are caveats: the source components may be nearly
colinear, especially if they are knots in an AGN jet; or the
magnification itself may vary across the source.  So it is desirable
to have more resolved features.  Nevertheless, if a model reproduces
the image positions for a three-component source, it can reasonably
claim to have fitted the tensor magnification.  With four images
having four components each, \galax\ arguably provides the most
detailed magnification information of any galaxy lens.
\cite{2000ApJ...535..671T} implicitly fitted the tensor magnifications
in their models.  However, they did not inspect their free-form model
in greater detail to see if the second galaxy~X or other substructures
were buried in it.  It would be interesting to reconstruct and
post-process their published models in this way.  However, in this
paper we start from their VLBI image maps and remodel the lens using a
different free-form modeling method.

We use the {\em PixeLens\/} modeling code \citep{2004AJ....127.2604S}
which generates ensembles of pixelated 2D mass maps that (a)~exactly
reproduce the given image positions and, where known, time delays
between images, and (b)~are consistent with a prior.  The data in
question are the image positions.  The prior requires (i)~the mass
pixels to be non-negative; (ii)~density gradient anywhere to point
within $45^\circ$ of the lens center; (iii)~any mass pixel to have at
most twice the average density of its neighbors, except for the
central pixel which is unlimited; (iv)~the circularly averaged mass
profile to be steeper than $R^{-0.5}$. The lens center is assumed to
coincide with the galaxy light peak, but no other information about
the galaxy light is supplied.  This prior is designed to incorporate
some conservative assumptions about the lenses. The earlier paper
includes more justification of the prior, while \cite{2006ApJ...650L..17S}
tests the prior against a galaxy-formation simulation.

\begin{figure}
\epsscale{0.5}
\plotone{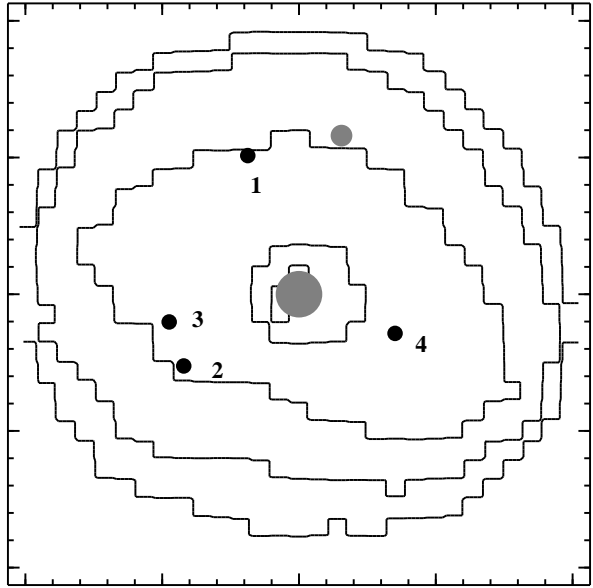}
\caption{Ensemble average mass map for \galax.  On the axes, the large
ticks show 10~kpc.  The mass contours are in powers of $10^{0.4}$
(i.e., like a magnitude scale) and the third outermost contour
corresponds to $\munit$.  Images are marked by small filled circles
and labelled by arrival time order; 1,2,3,4 correspond to B,A1,A2,C in older
papers.  In fact each filled circle really corresponds to three very
close image components; these are the VLBI components p,r,s shown in
Fig.~2 of \cite{2000ApJ...535..671T}.  The main lensing galaxy G
and the secondary galaxy X are indicated by gray filled circles, whose
radius is proportional the cube root of the flux.}
\label{J0414-map}
\end{figure}

Fig.~\ref{J0414-map} shows our mass map for \galax, actually the
average of an ensemble of 400 mass models.\footnote{See the Appendix
for further details on the interpretation of model ensembles and
ensemble-averages.} The input data were the
positions of the image components p,r,s from Fig.~2 and Table~1 of
\cite{2000ApJ...535..671T}, along with lens and source redshifts and
the centroid of the main lensing galaxy G.
(A fourth component, q, is very close to p, and not considered separately
in our models.)  The mass map shows an
asymmetry with more mass in the proximity of the second galaxy
X. Since X is only $\sim10\%$ as bright as G and the models had no
input on X, it would be very interesting if the models infer
substructure associated with X.  But before going on to substructure,
let us examine the models some more.

\begin{figure}
\epsscale{0.5}
\plotone{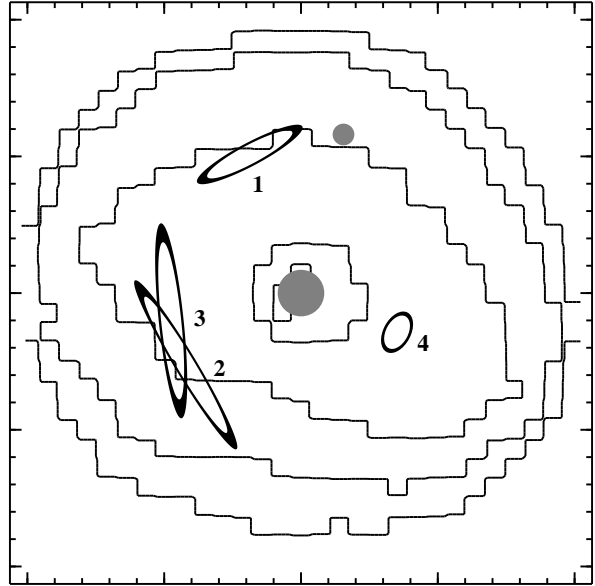}
\caption{Similar to Fig.~\ref{J0414-map}, but showing the inferred
tensor magnifications.  The ellipses indicate how a small
circular source would be magnified and distorted (except that the
sizes are greatly exaggerated).  This figure may be compared with
Fig.~3 in \cite{2000ApJ...535..671T}.}
\label{J0414-mag}
\end{figure}

In Fig.~\ref{J0414-mag} we show the model tensor magnifications with
ellipses.  Analogous ellipses appear in Fig.~3 of
\cite{2000ApJ...535..671T}, and from visual comparison the results
seem consistent.  The ellipses are oriented nearly radial-tangential
with respect to the center of G, but not quite: for image~2 the
misalignment is $\sim1^\circ$, while for the other images it is
5--7$^\circ$. The orientations vary only $\sim1^\circ$ across the
model ensemble, despite considerable variation in the mass maps
themselves.  Evidently the orientation of the magnification is
strongly constrained by the VLBI positions.

Now consider specifically inverse radial magnification $1-\psi_{rr}$,
where $\psi$ is the lens potential.  In the ensemble-average model
this quantity is in the ratio $0.95 \!:\! 1 \!:\!  0.88 \!:\! 0.78$
for images 1 through 4.  The ratios vary across the ensemble, but
images 2--4 consistently have their inverse radial magnifications in
decreasing order.  This is of interest because the generalized
isothermal potentials \citep{2000ApJ...544...98W} widely used in
parametric models of galaxy lenses predict that the radial
magnification should be equal for all images.  Could the total lens
potential here consist of a generalized isothermal for the main lens,
plus a small perturbation from galaxy X?  If so we would expect images
2--4, which are far from X, to all have equal radial magnification,
which is not the case.  So, it appears that even without X the lens is
not a generalized isothermal.  If galaxy-lens potentials are in
general significantly different from generalized isothermals, then
conclusions derived using this assumption
\citep{2005MNRAS.364.1459C,2005ApJ...626...51Y,2006ApJ...642...22Y}
may need to be revised.  A useful test would be to see whether a
generalized isothermal plus a perturbation from X can fit the VLBI
image structure in \galax.

\begin{figure}
\epsscale{0.6}
\plotone{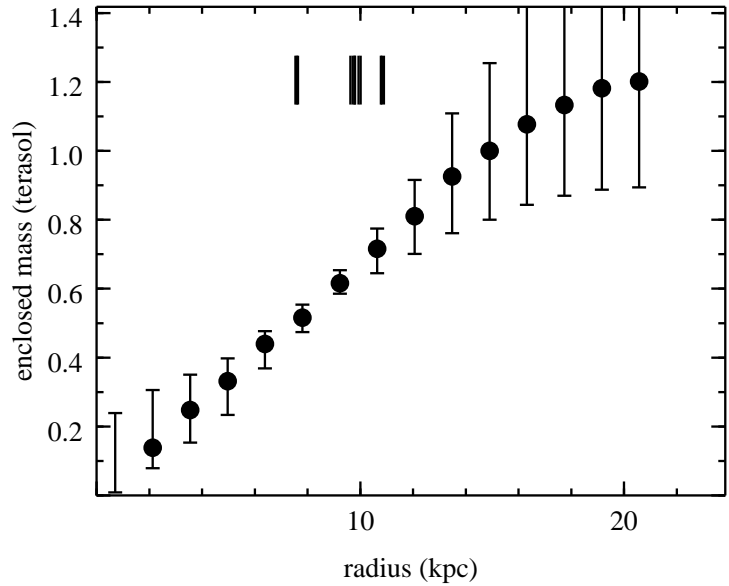}
\caption{Enclosed mass of \galax\ in $10^{12}M_\odot$.  The error
bars show a 90\% confidence region. The barcode-like pattern indicates
the image radii.}
\label{J0414-menc}
\end{figure}

Fig.~\ref{J0414-menc} plots the enclosed mass against projected
radius. We see that the total mass within the region of the images is
well-constrained, but the data allow the mass profile to be shallow or
steep.  This is simply an illustration of the well-known steepness
degeneracy that afflicts all lens models from a single source
redshift.  \citep[See e.g.,][for a recent discussion of degeneracies.]
{2006ApJ...653..936S}

\begin{figure}
\epsscale{0.5}
\plotone{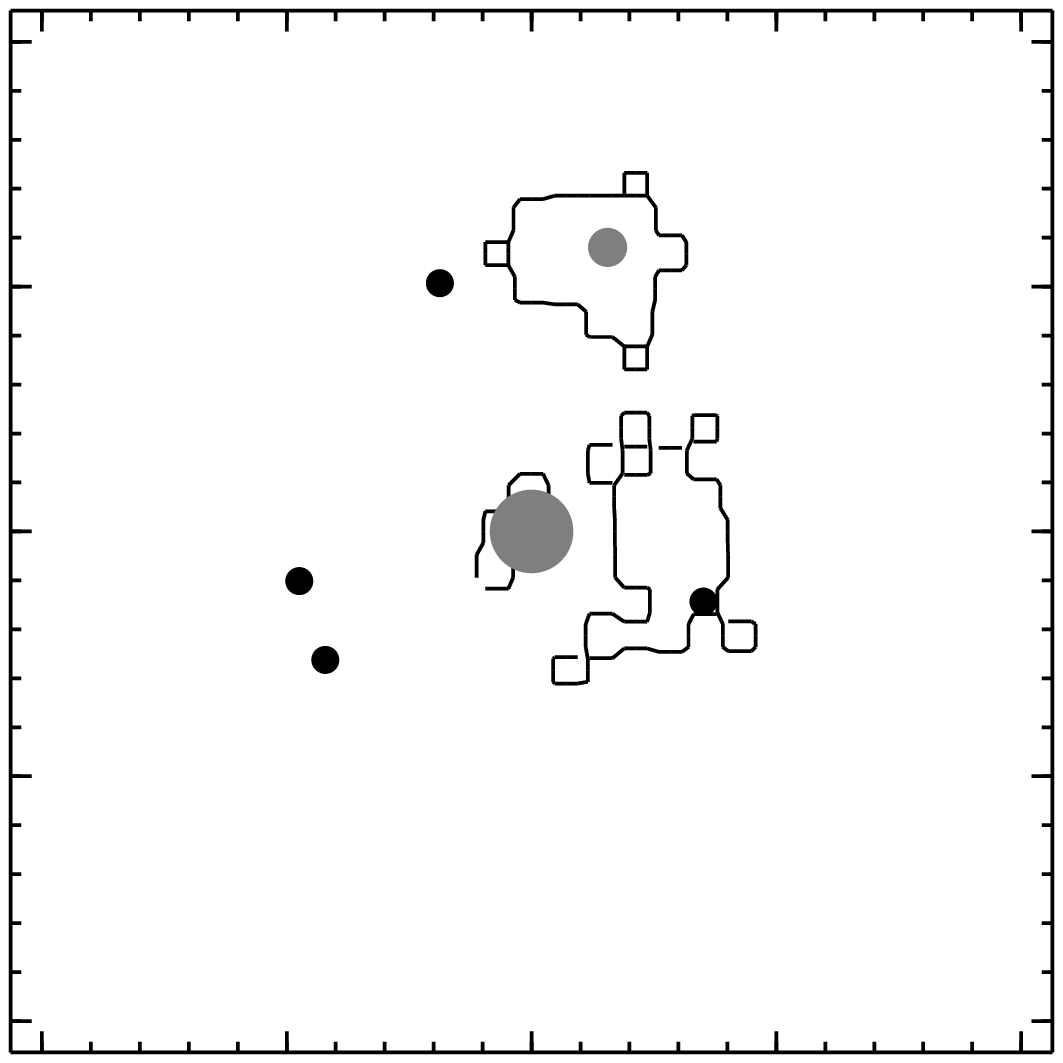}
\plotone{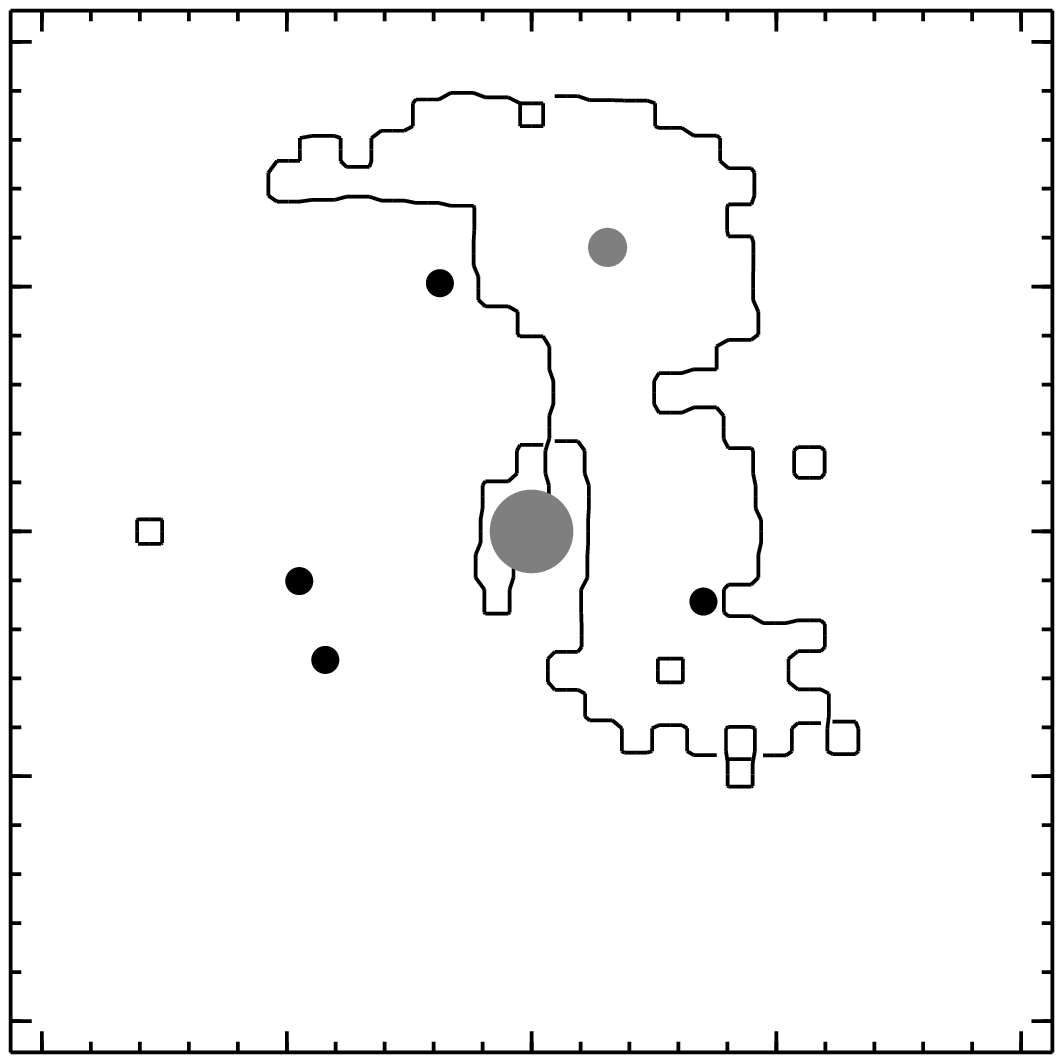}
\caption{Like Fig.~\ref{J0414-map} but showing the asymmetric
overdensity (see text) of \galax.  The contours in the upper
(lower) panel correspond to $0.2\;(0.1)\times\munit$ and contain an
excess mass of $\simeq1.7\;(2.5)\times10^{11}M_\odot$.
The total mass (cf.~Fig.~\ref{J0414-menc}) is $1.2\times10^{12}M_\odot$.}
\label{J0414-r}
\end{figure}

Because of the uncertainty in the mass profile, it is difficult to
define a base profile that can be subtracted off to isolate the
substructure.  We therefore define a simple residual,
\begin{equation}
\Sigma(\btheta) - \min\left[\Sigma(\btheta),\Sigma(-\btheta)\right].
\label{asymod}
\end{equation}
Here the second term is the maximal inversion-symmetric profile that
is non-negative everywhere.  The whole expression may be called the
asymmetric overdensity.  It is shown in Fig.~\ref{J0414-r}.  The
apparent coincidence of one overdensity with the secondary galaxy
X is intriguing.  Are we seeing part of a tidal tail?

We estimated the stellar mass in galaxies G and X using the observed
magnitudes and colors from CASTLeS\footnote{{\tt
http://www.cfa.harvard.edu/glensdata/}} and standard
population-evolution models, using the method described in
\cite{2005ApJ...623L...5F}, which yields $\sim5\times 10^{11}M_\odot$
for G and $\sim5\times 10^{10}M_\odot$ for X.  The former is roughly
half the total lensing mass, while the latter is roughly half the mass of the
substructure.  The interpretation of X as a satellite galaxy merging
into the halo of G seems plausible.  But it must be considered tentative.

\section{The cluster SDSS~\snclus}

This cluster initially attracted attention as a host of a giant quadruple quasar
\citep{2003Natur.426..810I,2004ApJ...605...78O}.  Subsequent data have
been greatly enhanced by the discovery of a central image of the
quasar \citep{2005PASJ...57L...7I}, a time-delay measurement
\citep{fohlmeister} and most importantly the discovery of three new
multiply-imaged galaxies \citep{2005ApJ...629L..73S}.  The latter work
provides 9 new images with a range of redshifts and covering a larger
area than the original quasar images; it also develops a mass model
with a slowly varying smooth part and discrete components associated
with galaxies, which fits all the image data.

An early attempt to map the substructure in this cluster was made in
\cite{2004AJ....128.2631W} using just the four quasar-image positions
then available.  The results varied with the priors used (four priors
were illustrated) and the conclusions relied on indirect arguments.
The paper argued for a shallow mass profile with up to 10\% of the
mass in substructure, with the substructure spatially correlated with
galaxies.  In hindsight these conclusions appear broadly correct.  The
paper also favored models where the short time-delay was $\leq20\,$d
and the external mass had a roughly E/W long axis over models with
longer time delays and/or a roughly N/S axis for the external mass,
but these conclusions were incorrect---the short time delay for the
quasar is measured as $38.3\pm2\,$d, while the quintuply-imaged galaxy
shares the typical morphology of short-axis quads
\citep{2003AJ....125.2769S} indicating the external mass axis to be
roughly NWN or SES.

The problems encountered in \cite{2004AJ....128.2631W} largely
disappear once we include the new multiple-imaged systems from
\cite{2005ApJ...629L..73S}.  \cite[][; hereafter Paper~I]{2006ApJ...652L...5S}
presents an ensemble of models fitting 13 images from the four known
sources.  (One of the two central images is omitted because of
resolution limits in the code.)  The technique is the same as we used
above for \galax, except for two important differences:
\begin{enumerate}
\item The minimal steepness requirement of $R^{-0.5}$ is dropped,
since for clusters a shallower profile is expected than for galaxies.
\item The range in $\zs$ is incorporated.  The availability of a range
of $\zs$ breaks the steepness degeneracy \citep{1998AJ....116.1541A}
seen in Fig.~\ref{J0414-menc} above and which plagues galaxy-lensing
generally.
\end{enumerate}
The prior is weaker than any of the priors in
\cite{2004AJ....128.2631W}, yet the mass profile and the mass map in
general are much better constrained.  Paper~I concentrates on the
circularly-averaged mass profile, finding it to agree with the inner
profiles typical of cold dark matter simulations, postponing the study
of substructure.  But we remark that in the models of \snclus\ in
Paper~I, the short time delay ranges from 0 to 60\thinspace d (in
other words, it is not well-constrained but is consistent with the
measured value), while the fitted external shear implies the long axis
of external mass to be NWN/SES.

\begin{figure}
\epsscale{0.5}
\plotone{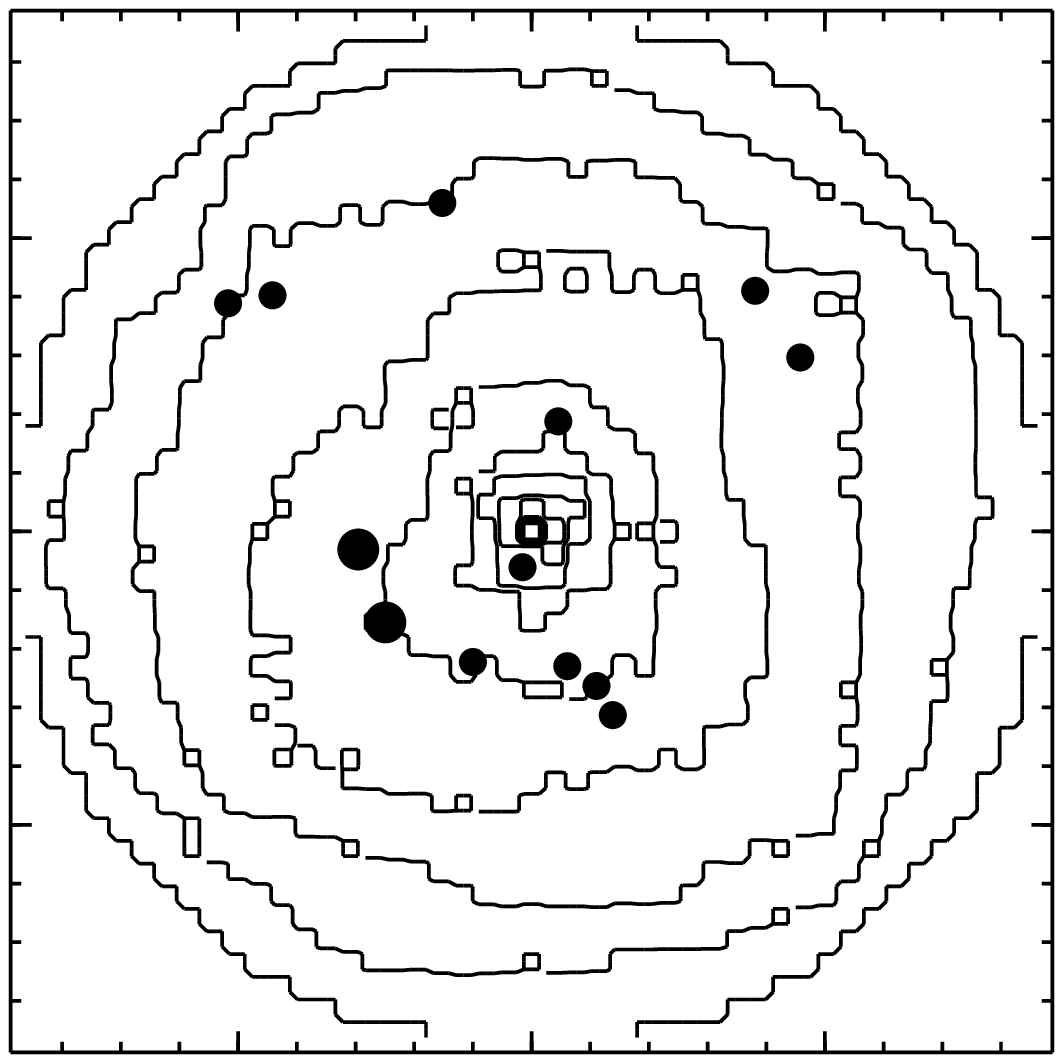}
\caption{Ensemble-average mass map for \snclus.  Large ticks on the
axes correspond to $100\kpc$.  Contour steps are $0.5\times\munit$.
Filled circles mark image positions.  This figure is similar to the
top left panel in Fig.~1 of \cite{2006ApJ...652L...5S}, except that the short
time delay (two larger circles to the SE) is constrained to be
38\thinspace days.  Fig.~9 of \cite{fohlmeister} also fits the time
delay, but includes only one of the four multiple-image systems.}
\label{J1004d-mq}
\end{figure}

Adding the time-delay as a constraint is straightforward. In
Fig.~\ref{J1004d-mq} we show the ensemble-average of 400 models
fitting the measured time delay as well as the 13 images from four
sources.  Comparing with Fig.~1 in Paper~I shows only very small
differences.  In other words, once the three additional image-systems
have been incorporated, the short time-delay makes only a small
incremental difference.  The \cite{fohlmeister} measurement is
nevertheless important, because it sets a baseline for future
measurements of the longer delays, which will provide stronger
constraints on the mass distribution and possibly an accurate
single-lens measurement of $H_0$.

\begin{figure}
\epsscale{0.5}
\plotone{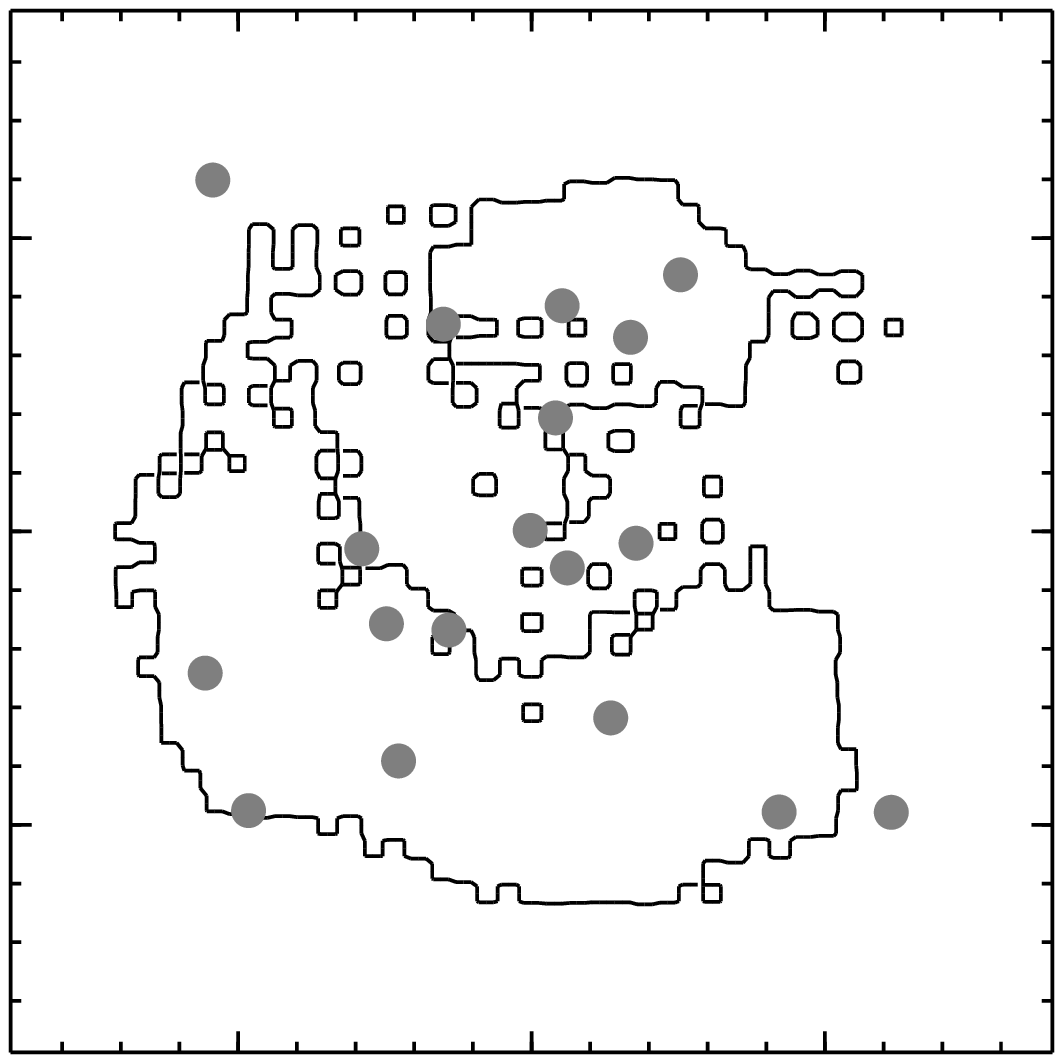}
\plotone{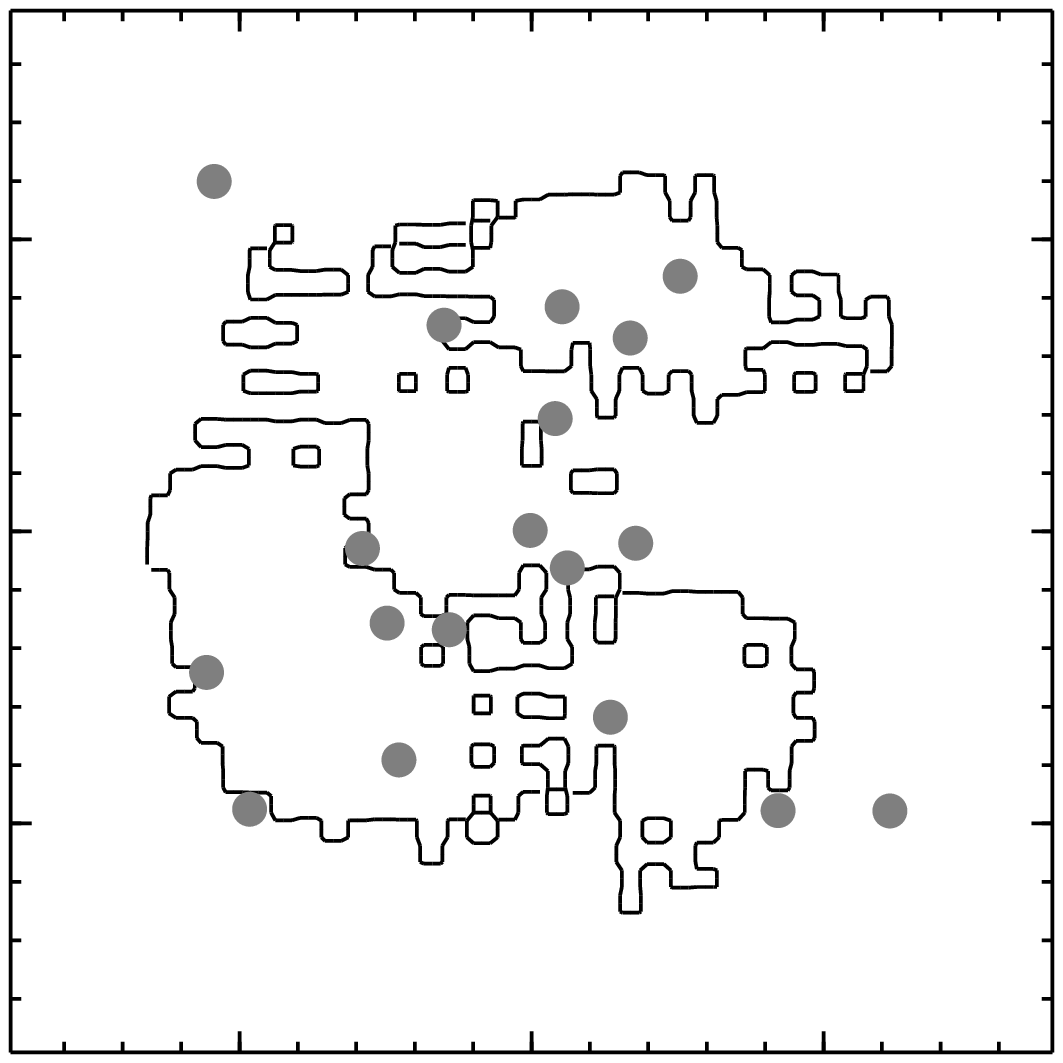}
\caption{Residual mass maps of \snclus\ after substracting off the
best-fit NFW.  The upper panel is derived from the model shown in
Fig.~\ref{J1004d-mq}, while the lower panel is derived from the models
in Paper~I. The contours show an overdensity level of $10^8M_\odot$
and enclose $\simeq7\,(6)\times10^{12}M_\odot$ of excess mass in the
upper (lower) panel.  This substructure is $\simeq7\%\,(6\%)$ of the
total mass in the field. As in Fig.~\ref{J1004d-mq}, large ticks on the
axis show 100~kpc.  Gray filled circles indicate early-type galaxies.}
\label{J1004-rg}
\end{figure}

Proceeding now to substructure, we subtract off the best-fit NFW
profile from the mass map.  That is to say, we fit the projected form
\citep{1996A&A...313..697B} of the spherical
NFW profile to Fig.~\ref{J1004d-mq}
and subtract it off.  One can imagine other, perhaps better,
procedures for uncovering substructure, but this one seems the
simplest.  We do the same with the mass map of \snclus\ from Paper~I,
which lacks the time-delay constraint and uses slightly larger pixels.
Fig.~\ref{J1004-rg} shows the results.\footnote{Recall that
Fig.~\ref{J1004-rg} refers to the average of an ensemble of 400
models.  See Fig.~\ref{J1004-sub} for results from smaller ensembles,
down to a single model.} We see overdensity mostly to
the N and S.  This is broadly consistent with the substructure maps
over a smaller area in \cite{2004AJ....128.2631W}, which used a
different method to subtract off the smooth portion.
Fig.~\ref{J1004-rg} overplots the probable cluster galaxies.  These
are galaxies brighter than $i=24$, with a color cut following
\cite{2004ApJ...605...78O}.  From Fig.~10 of that paper, we estimate
the typical galaxy to have $M_r\simeq-20.5$ or $\sim10^{10}L_\odot$.
We have not attempted a population-evolution model for the stellar
mass, but the latter is probably less than the mass in the
substructure (6--7$\times10^{12}M_\odot$) and can be only a few
percent of the total in the field.

Fig.~\ref{J1004-rg} unfortunately also shows an artifact common in
{\em PixeLens\/} mass maps.  Although the density-gradient constraint
in the prior (see Section~\ref{sys1}) disallows mass peaks, the
discretized form that is actually implemented disallows extended hills
but allows one-pixel peaks.  Usually this artifact is only noticeable
as single-pixel blemishes \citep{1997MNRAS.292..148S} but subtracting
off a profile one-pixel peaks become proportionately higher.  In
particular, Fig.~\ref{J1004-rg} shows a corrugation effect in some
places, consisting of an array of one-pixel peaks.

Are the galaxies correlated with the substructure?  It is easy to make
a statistical test for this \citep[an improved version of the test
used in][]{2004AJ....128.2631W} as follows.  Take the total lens mass
within a projected 
distance $R$ of each galaxy, and sum over all galaxies (weighted by
the galaxy brightness if available).  Call this $h(R)$.  Next,
calculate $\tilde h(R)$ by rotating each galaxy around the cluster
center by a random angle.  Let $S(R)$ be the percentage of cases where
$\tilde h(R)<h(R)$.  We may call this the percentage
significance.\footnote{Percentage significance is not a statistics
term.  What we call ``95\% significant'' is known in statistics as
``$p$~value of 0.05'', and will be significant (or not) if it is less
(or more) than the pre-specified significance {\em level}. But apart
from the use of astronomers' vernacular, our procedure is standard
statistics.}  If $S(R)$ is close to 100\% then the galaxies are
correlated with non-radial variations in the density, that is, with
substructure.

\begin{figure}
\epsscale{0.5}
\plotone{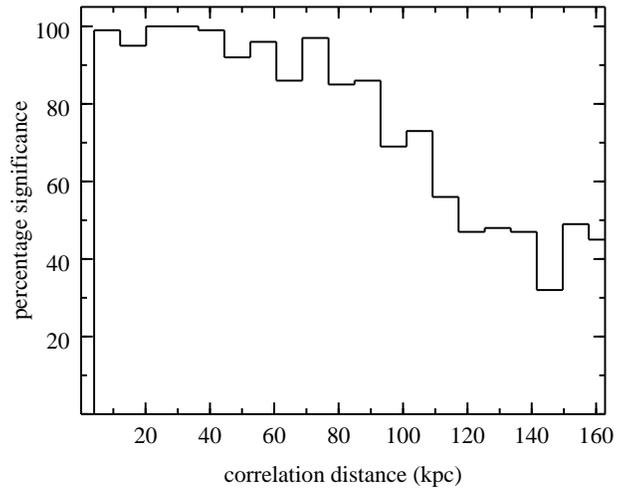}
\caption{Correlation significance $S(R)$ for galaxies against
substructure in \snclus.  This refers to the models including the time
delay (i.e., Fig.~\ref{J1004d-mq} and the upper panel of
Fig.~\ref{J1004-rg}) but the answer for the other case is
qualitatively similar.}
\label{J1004-corr}
\end{figure}

Fig.~\ref{J1004-corr} shows that galaxies are correlated with
substructure to scales of $\sim50\kpc$.  (Short range correlation is
not significant because the mass reconstruction is noisy on the
pixel-to-pixel scale.)  Note that the correlation does not mean that
most of the galaxies are associated with overdensities, it only
implies that overdensity increases the chance there being a galaxy,
even $\sim100\kpc$ away.  The overdensities must hence be interpreted
as genuine extended structures, not statistical associations of much
smaller structures.  Hence the term {\em meso-structure}.

Now the total cluster density at $r\sim100\kpc$ is
$\simeq4\times10^6M_\odot\kpc^{-3}$ (cf.~Paper~I).  This gives a
crossing time $(G\rho)^{-\frac12}\simeq2\times10^8\,\rm yr$,
much shorter than the
age of the cluster.  Hence meso-structure of $\sim10\%$ of the total
mass on a scale of $\sim100\kpc$ would have gotten mixed long ago.
Thus the observed meso-structure seems most plausible as either a
line-of-sight filament or as a merger in progress.

\section{The cluster \abclus}

We saw above that the observations of \snclus\ allowed a more detailed
and confident mapping of substructure.  For \abclus\ the situation is
much better still: the observations \citep{2005ApJ...621...53B} are
enough to reconstruct the lens twice from two different data sets;
additionally, redshifts of individual cluster galaxies are available.
As well as \cite{2005ApJ...621...53B}, several other groups have
modeled these data
\citep{2005MNRAS.362.1247D,2006ApJ...640..639Z,2006MNRAS.372.1425H,
2006astro.ph.12165L}.
These papers
all agree that the cluster is compatible with an NFW profile plus some
substructure, but the substructure itself has not been discussed in
detail, apart from a distinctive clump to the NE. (This clump is
oriented towards 2~o'clock in some published figures.)

\begin{figure}
\epsscale{0.5}
\plotone{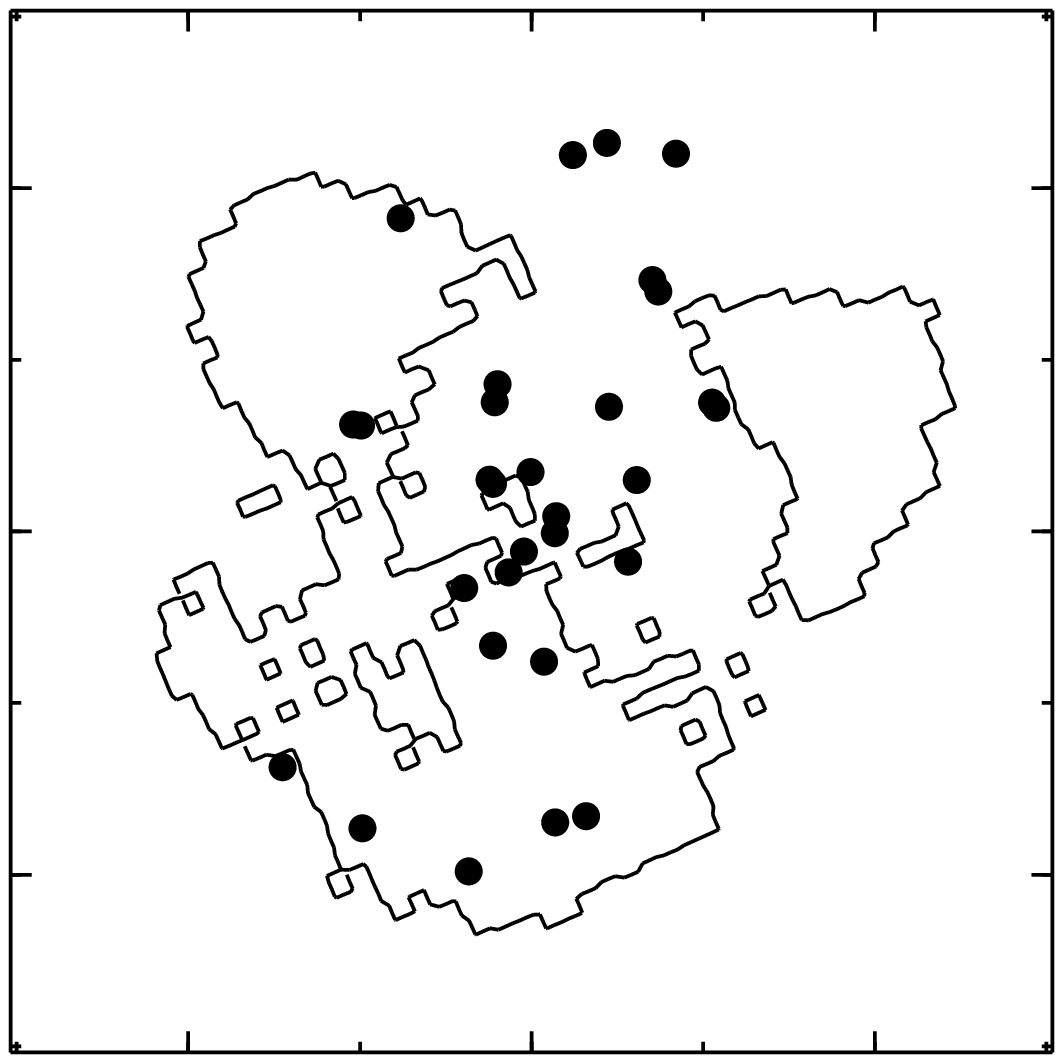}
\plotone{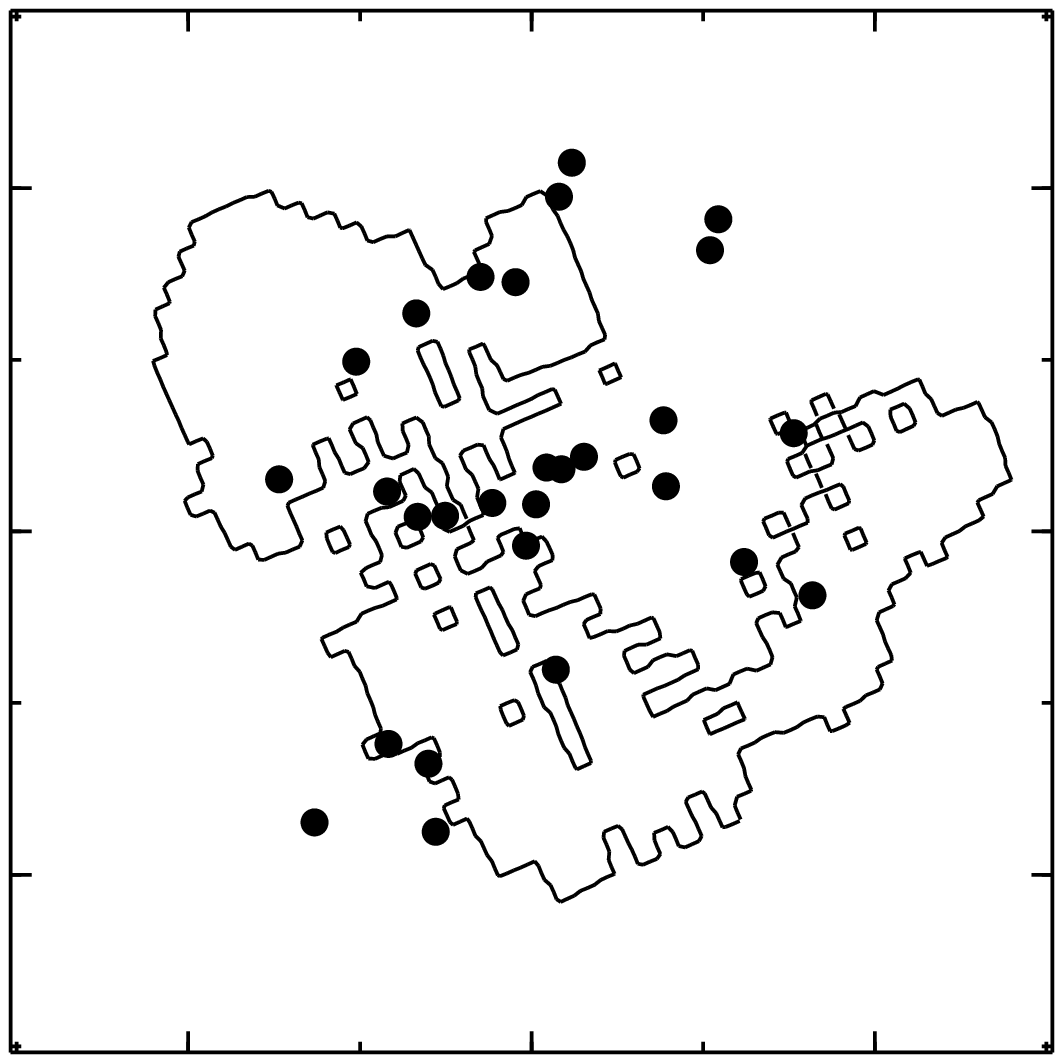}
\caption{Residual mass maps of \abclus, using two sets of lensed
images (filled circles).  Large ticks on the axis show 100~kpc. In both
panels, the contour denotes an overdensity level of
$2\times10^8M_\odot$ and encloses $\simeq3\times10^{13}M_\odot$ of
excess mass.  This substructure amounts to $\simeq7\%$ of the
total mass in the field.
(This figure has the standard N-up/E-left orientation,
but the mass pixel array follows the coordinate axes from Table~2 of
\cite{2005ApJ...621...53B} and these coordinate axes are discernable
from the contours.)}
\label{A1689-rq}
\end{figure}

Mass maps of the cluster using two sets of multiple images and using
the same prior as for \snclus\ are presented in Paper~I.
Fig.~\ref{A1689-rq} shows the result of substracting off the best-fit
NFW profile from these mass maps.  Although our initial reason for
modeling two subsets of the data that {\em PixeLens\/} currently
cannot handle all the data together, we have here an unexpected bonus: similar
substructure comes out of independent data sets.  The NE clump and two
other overdense regions are evident.  \cite{2005MNRAS.362.1247D}
mention three substructure clumps in passing, but we do not know if
they identified the same ones.

\begin{figure}
\epsscale{0.5}
\plotone{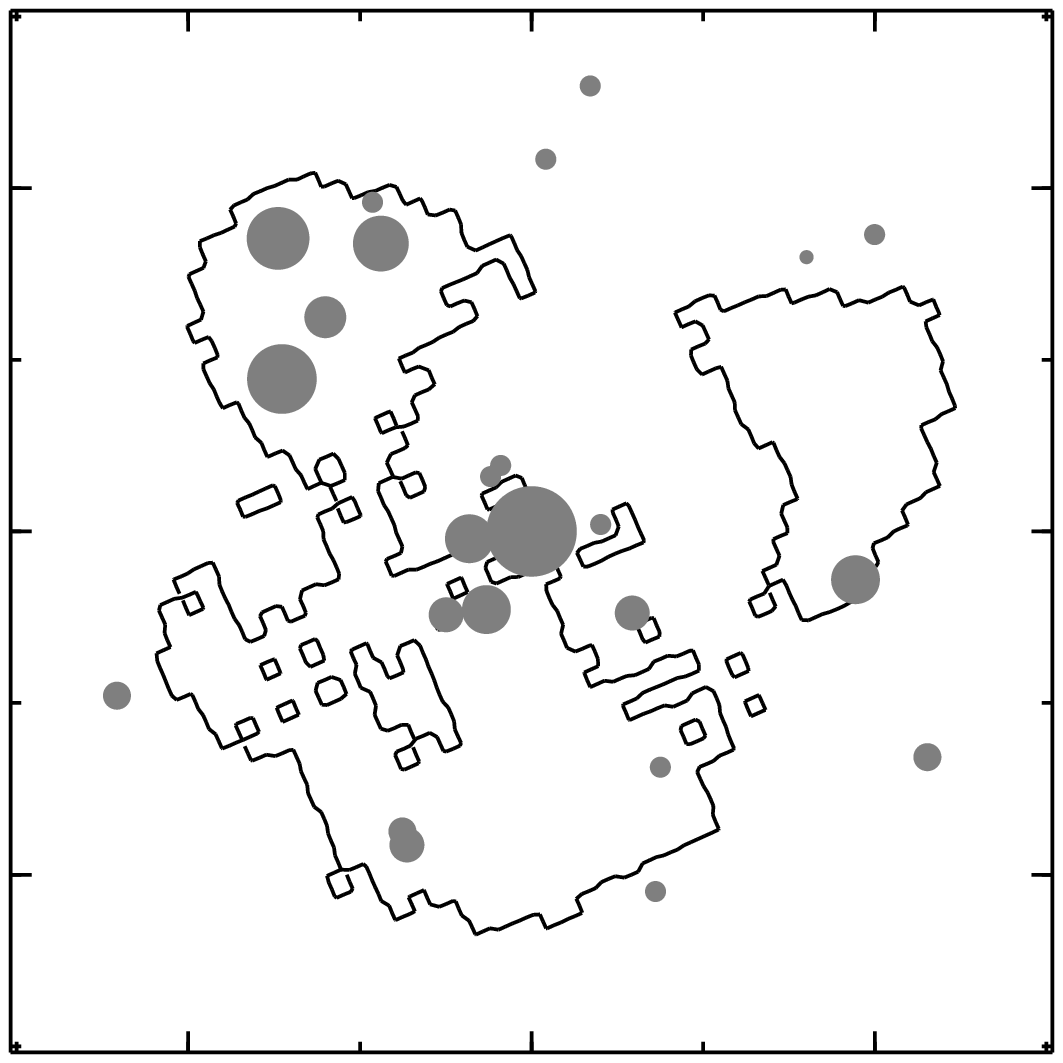}
\plotone{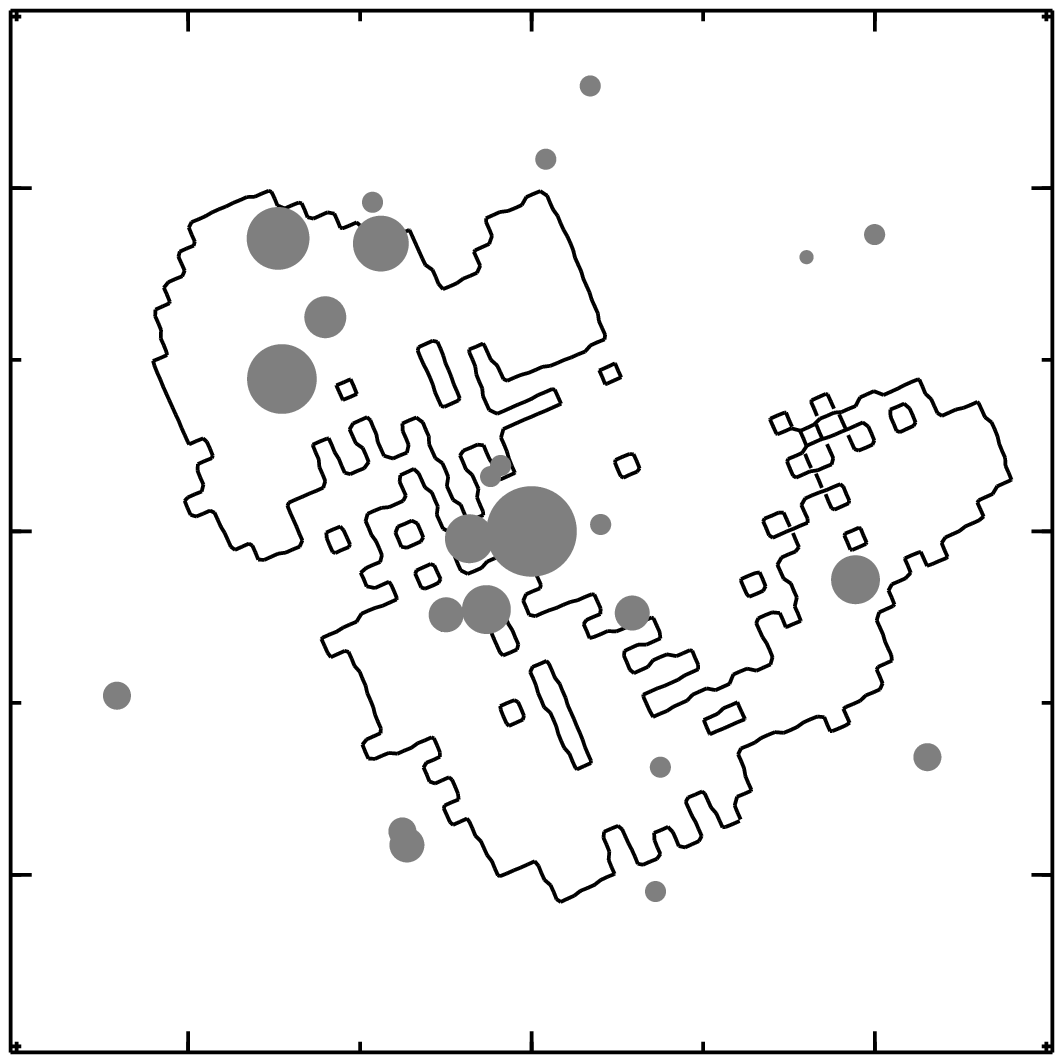}
\caption{Residual map of \abclus\ with probable cluster members
shown. The contours are as in Fig.~\ref{A1689-rq}.  Galaxies with
$B-R>2$ are indicated by gray circles, shown with radius proportional
to the cube root of the flux.}
\label{A1689-rg}
\end{figure}

\begin{figure}
\epsscale{0.5}
\plotone{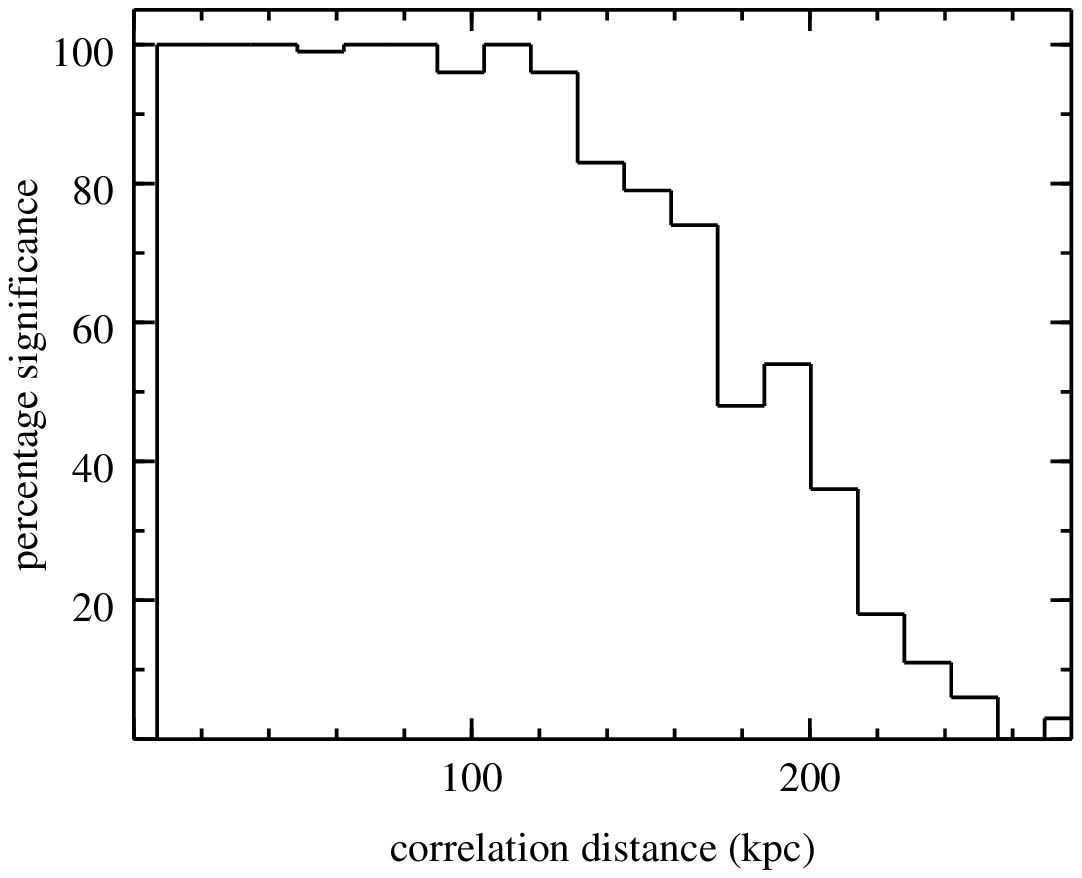}
\plotone{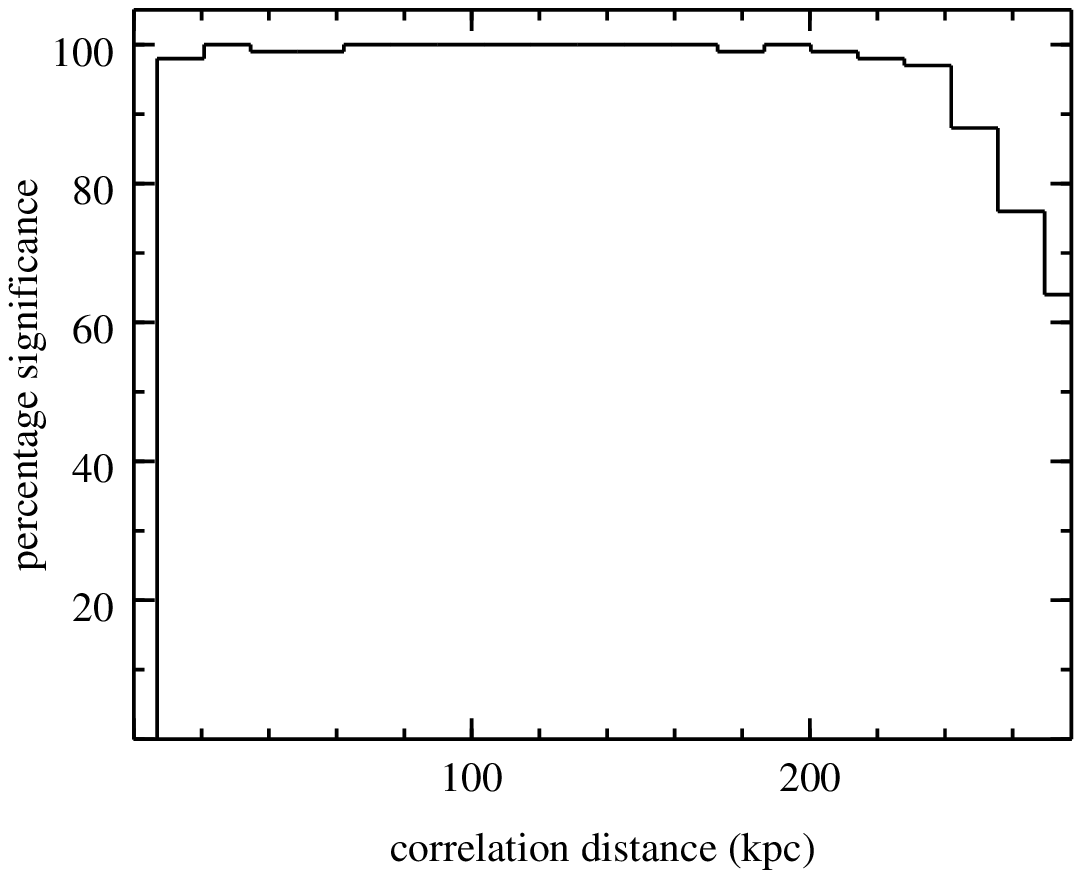}
\caption{Correlation significance for flux-weighted galaxies against
substructure ACO~1689. Panels correspond to panels in
Fig.~\ref{A1689-rg}.  The longer-range correlation seen in the
lower panel is presumably a manifestation of the isthmus to the SW in
the lower panel of Fig.~\ref{A1689-rg}.}
\label{A1689-corr}
\end{figure}

Colors and redshifts of galaxies in the field have been tabulated by
\cite{2002A&A...382...60D}.  A break around $B-R=2$ is noticeable in
the observed colors, and accordingly we suggest that galaxies redder
than this are probable cluster members.  Fig.~\ref{A1689-rg} shows
these galaxies.  We have not attempted a population-evolution model,
but a very rough extrapolation from \cite{2005ApJ...623L...5F}
indicates a few times $10^{12}M_\odot$ in stellar mass, much less than
the meso-structure.  The meso-structure does, however, correlate with
the galaxies, as Fig.~\ref{A1689-corr} shows.  As in \snclus, we
see correlations, but in this cluster they extend further,
to $\sim100\kpc$ scales or even beyond.

As with \snclus, the most plausible interpretation
of the $\sim10\%$ meso-structure is
again as either line-of-sight filaments or as ongoing mergers.

\begin{figure}
\epsscale{0.5}
\plotone{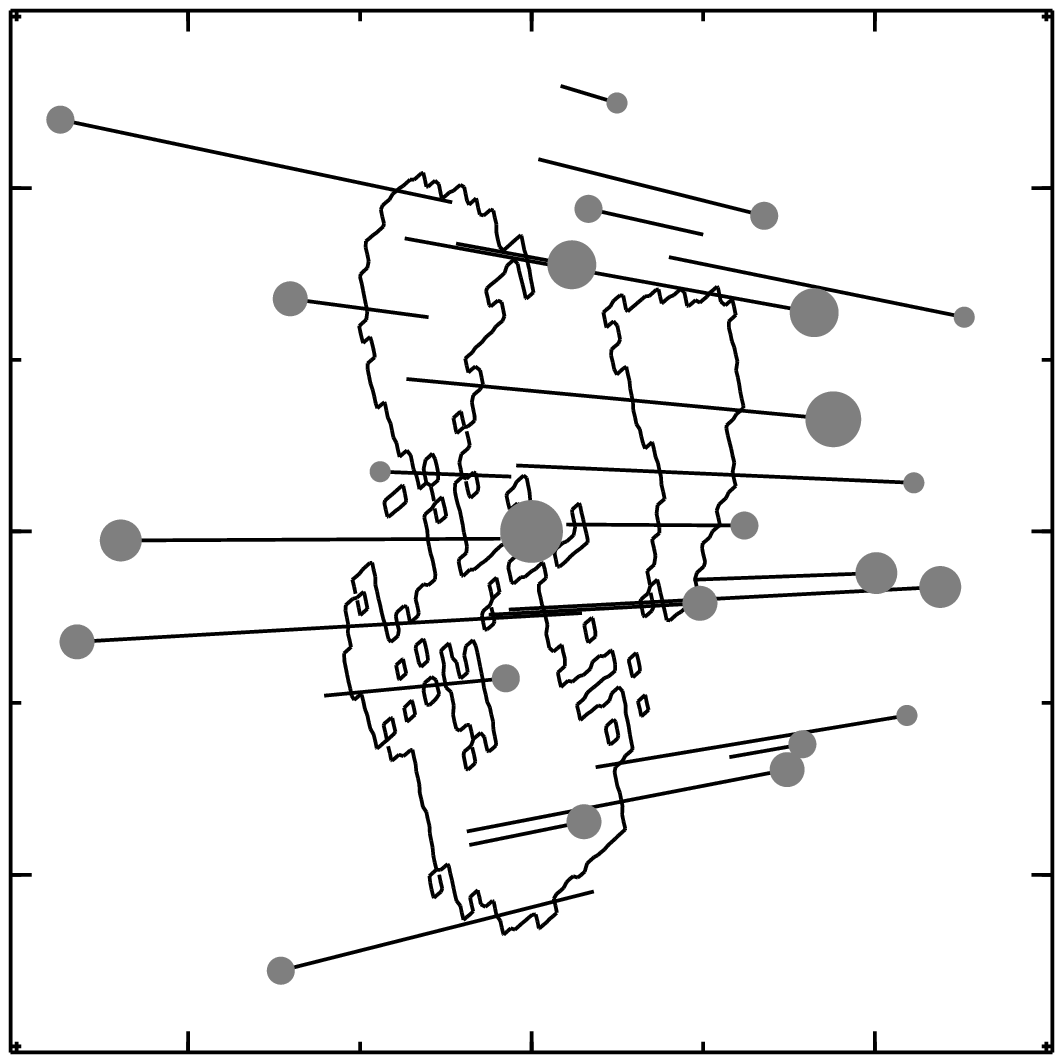}
\plotone{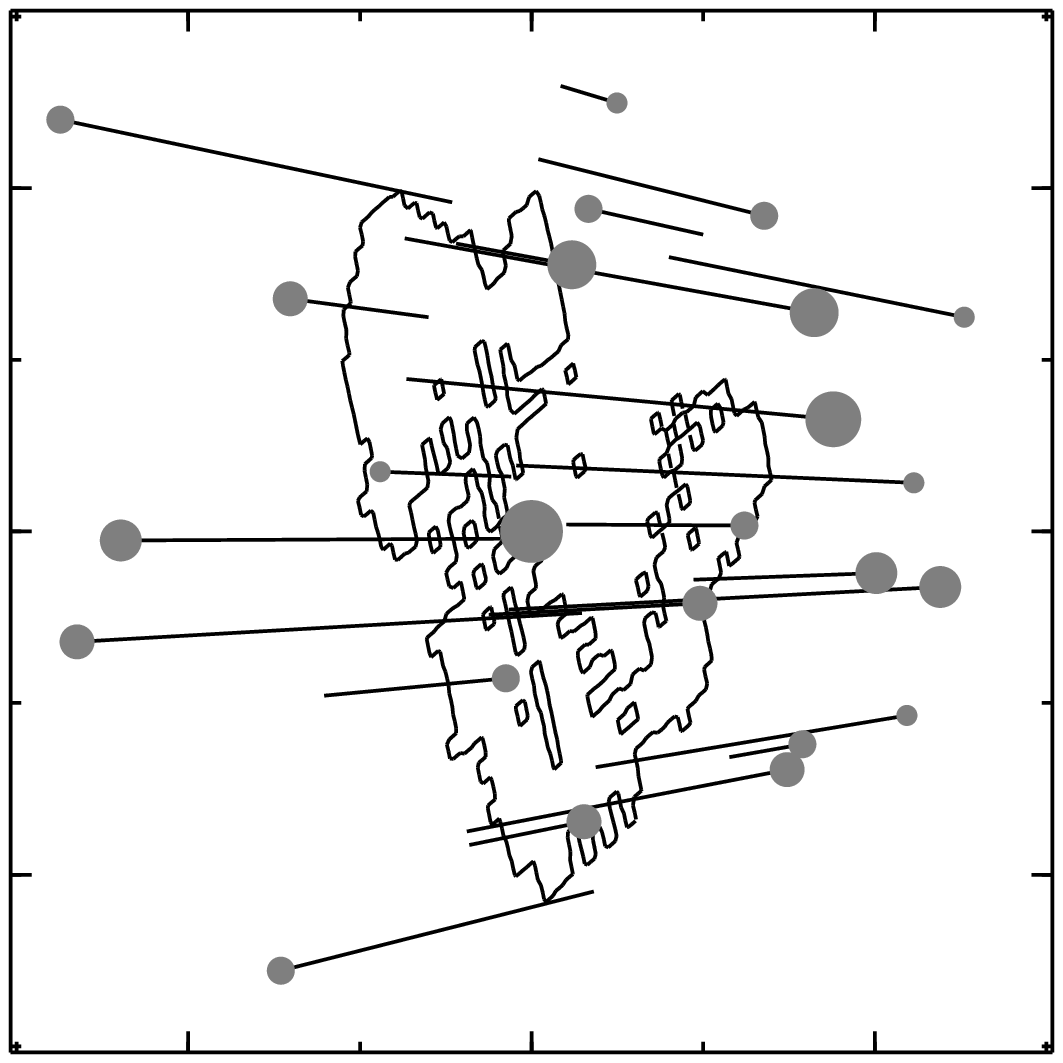}
\caption{A perspective version of Fig.~\ref{A1689-rg}, taking the
galaxy redshifts as the third coordinate. The thin lines indicate the
projection of each galaxy onto the notional lens plane, which is taken
to be coplanar with the brightest cluster galaxy.  The range is
$\sim500\kpc$ in the spatial direction, and $\sim10\,000\;\rm
km/sec$.}
\label{A1689-rgp}
\end{figure}

We make an attempt to test for these possibilities.  In
Fig.~\ref{A1689-rgp} we show a perspective version of
Fig.~\ref{A1689-rg} with the third dimension being the redshift
relative to the brightest cluster galaxy.  There are no distinctive
groups in this plot, so it does not appear that \abclus\ consists of
two or more isolated clusters.  This appears to favor the filament
interpretation.

\section{Discussion}

Our results leave no doubt that substructure at the 10\% level can be
mapped in the best-observed strong-lensing systems.  The inferred
substructure is tentative for the galaxy lens \galax, more confident
for the cluster lens \snclus, while for \abclus\ the substructure is
explicitly verified by using independent subsets of the data.  It
turns out that, although the link between magnification and
substructure has attracted much attention recently, in fact multiple
sources provide more information on substructure than magnification.

The substructure we find is meso-structure: extended features much
larger than galaxies and more massive than the stellar content, but
correlated with galaxies.  Such features would not survive unmixed in
the inner regions of the galaxies or clusters, and they are improbable
as chance alignments.  Evidently they are interacting or otherwise
dynamically evolving.

Much more can be done with current or near-future data, especially in
the case of \abclus.  One important goal is to make the lens
reconstructions finer grained.  Here progress using adaptive
resolution has been made
\citep{2001ASPC..237..279S,2005MNRAS.362.1247D} but the real challenge
is to resolve the halos of individual galaxies.  A study of
cluster-galaxy redshifts and the implied dynamics would help resolve
the 3D structure.  More understanding of the gas dynamics is needed,
not least the puzzle of why the equilibrium models for x-ray gas
underestimate the total mass \citep{2004ApJ...607..190A}.  These are
all messy problems, but worth research effort, because the variety of
data on \abclus\ makes it perhaps our most hopeful place for
understanding the physics of cluster and galaxy formation.

\acknowledgements We thank Paul Schechter for the appropriate mix of
encouragement and criticism, and for suggesting the term
`meso-structure'.

\appendix

\section{Inside the model ensembles}

The results in this paper derive from ensembles of pixelated models.
(We have five such ensembles: one for \galax, two from slightly
different data on \snclus, and two using disjoint data sets on
\abclus.  Each ensemble has 400 models.)  Each model in an ensemble is
required to reproduce the data precisely; subject to this constraint,
the ensemble samples models according to the prior.

Such model ensembles are by now a standard technique in data analysis,
described in several books \citep[for example,][]{2003pda.book}, and
known by different names such as Metropolis algorithm or Markov-Chain
Monte-Carlo.  The ensemble has the interpretation of a Bayesian
posterior probability distribution for the model --- which in our case
is the mass map --- and from this two very useful properties follow.

\begin{enumerate}
\item A histogram of any quantity (for example, the enclosed mass
within $100\kpc$) across the ensemble is the posterior probability
distribution for that quantity, with all other variables marginalized
out.  Hence error bars, including asymmetric error bars, can be
trivially extracted from the ensemble.
\item The ensemble average of any quantity is its expectation value.
Thus the ensemble-average models we have illustrated have each pixel
set to its expectation value.
\end{enumerate}

The above properties apply also to other lens-ensemble methods
\citep{2003ApJ...590...39K,2004ApJ...605...78O}.  However, {\em
PixeLens\/} ensembles have a further useful property.  Since the data
constraints are expressed as linear equations and the prior is
formulated as linear inequalities, it follows that the
ensemble-average model automatically fits both data and prior.  The
ensemble-average model in {\em PixeLens\/} is thus a very useful one
to illustrate: it fits the data, satisfies the prior, and is also the
expectation value across the whole ensemble.

\begin{figure}
\epsscale{0.4}
\plotone{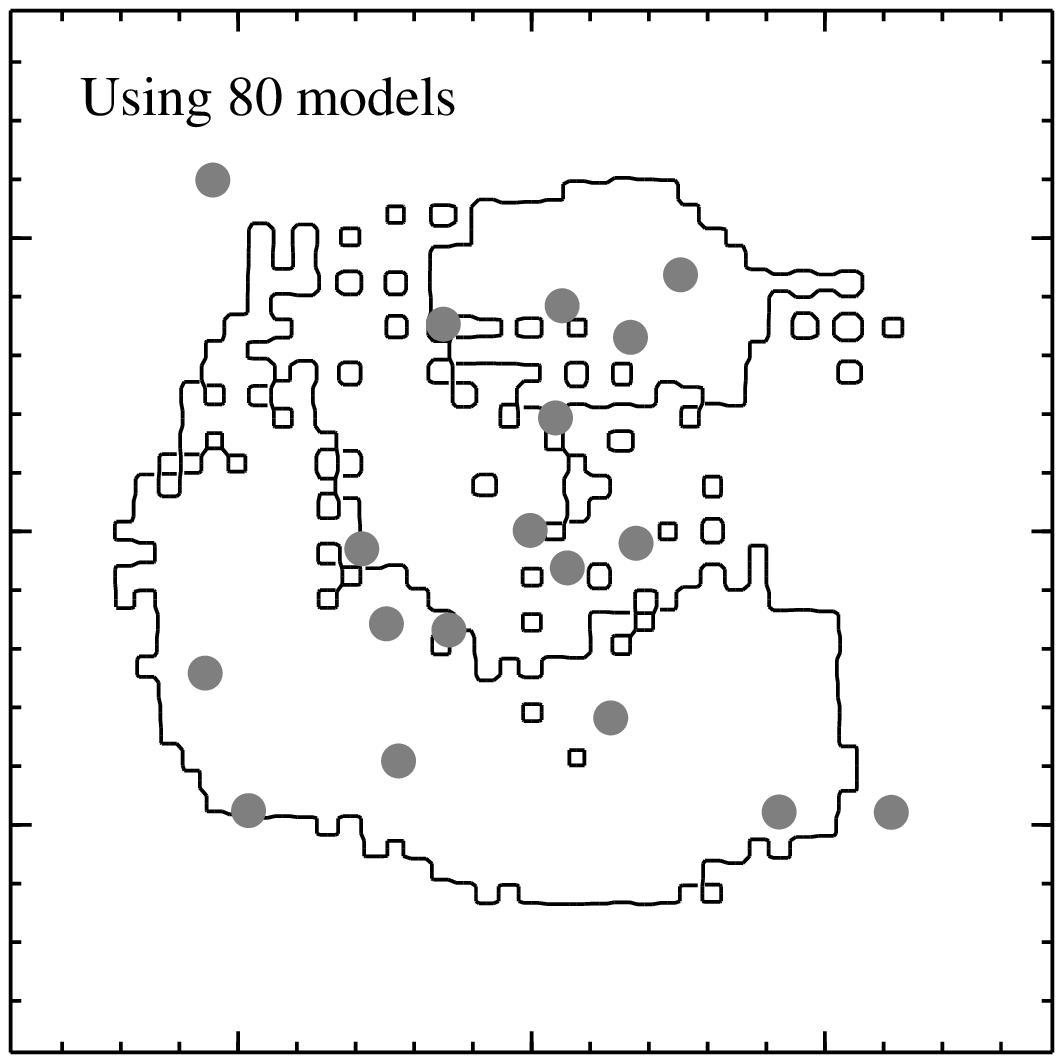}
\plotone{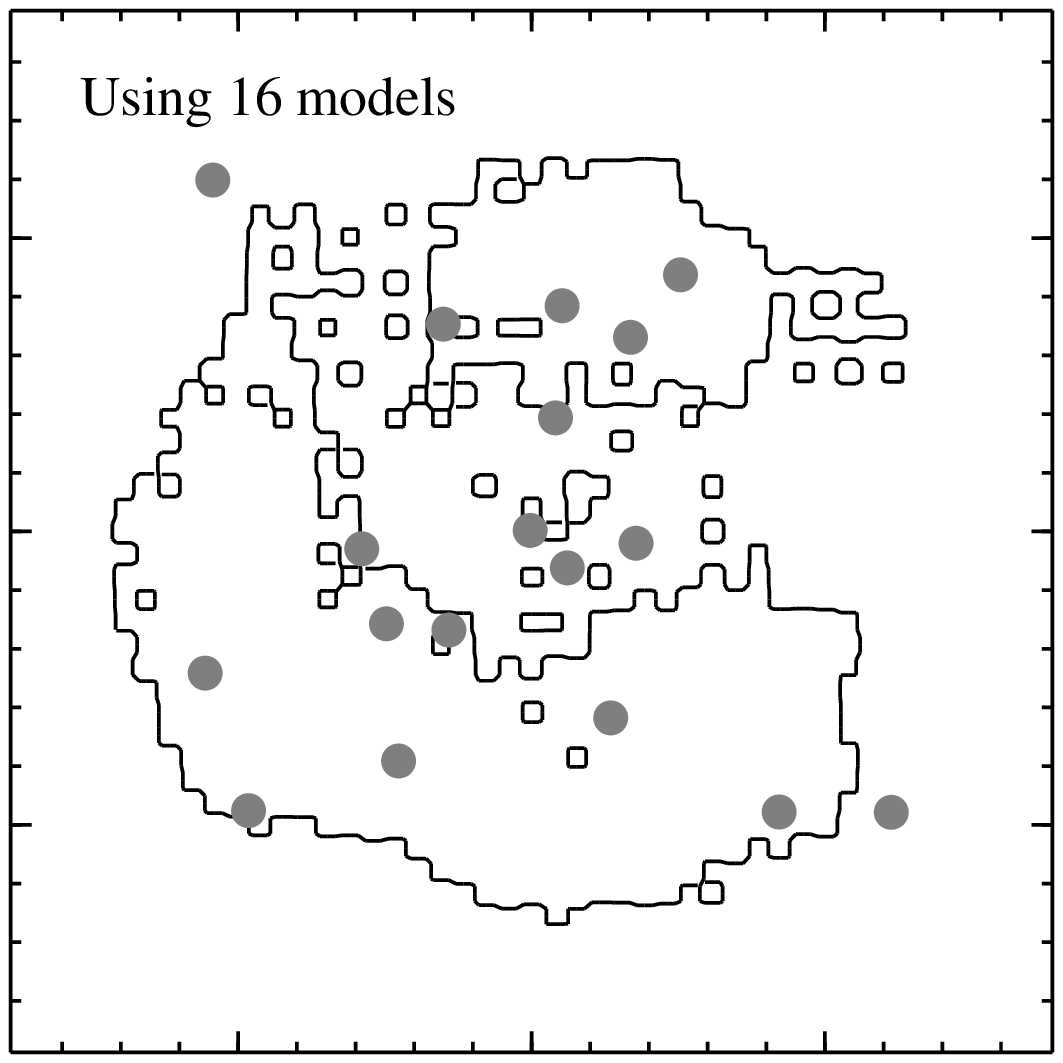} \goodbreak
\plotone{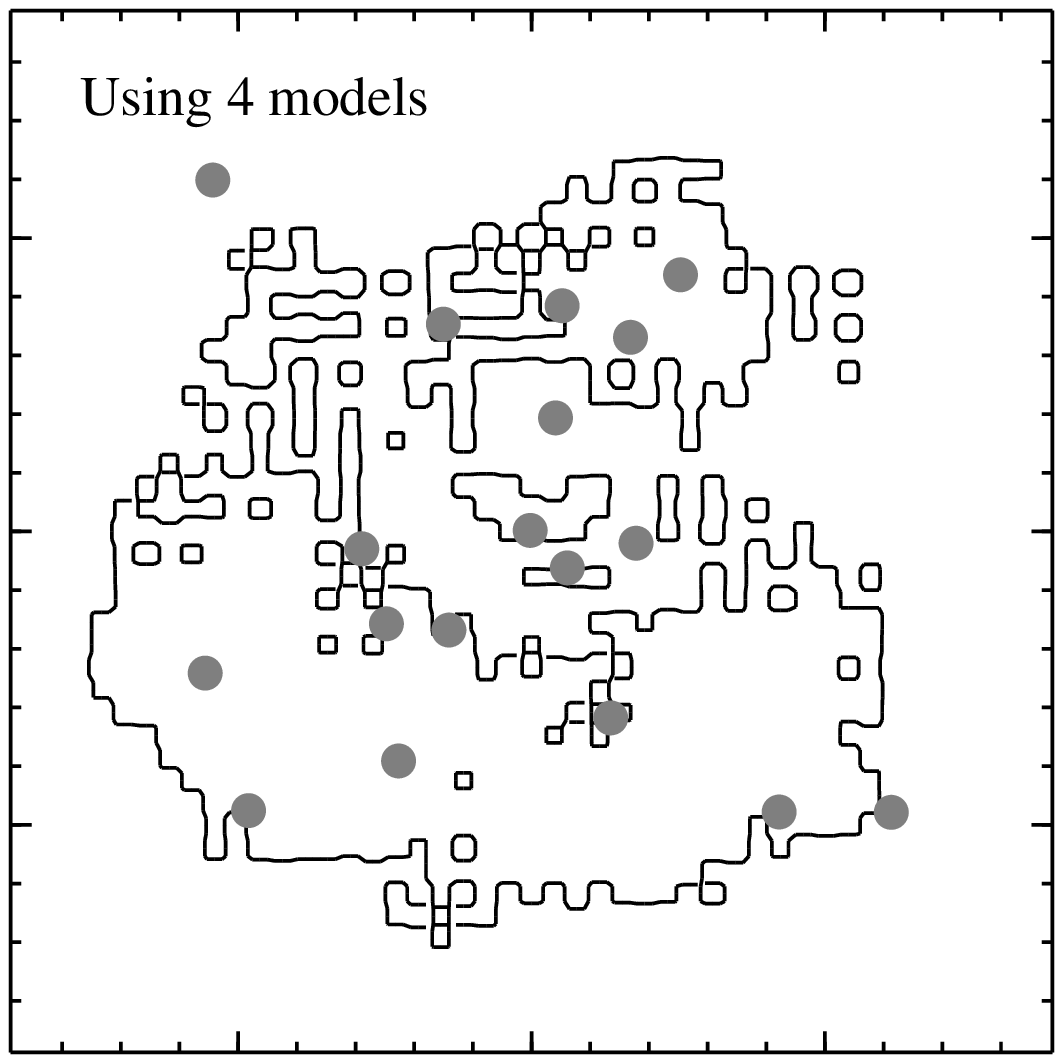}
\plotone{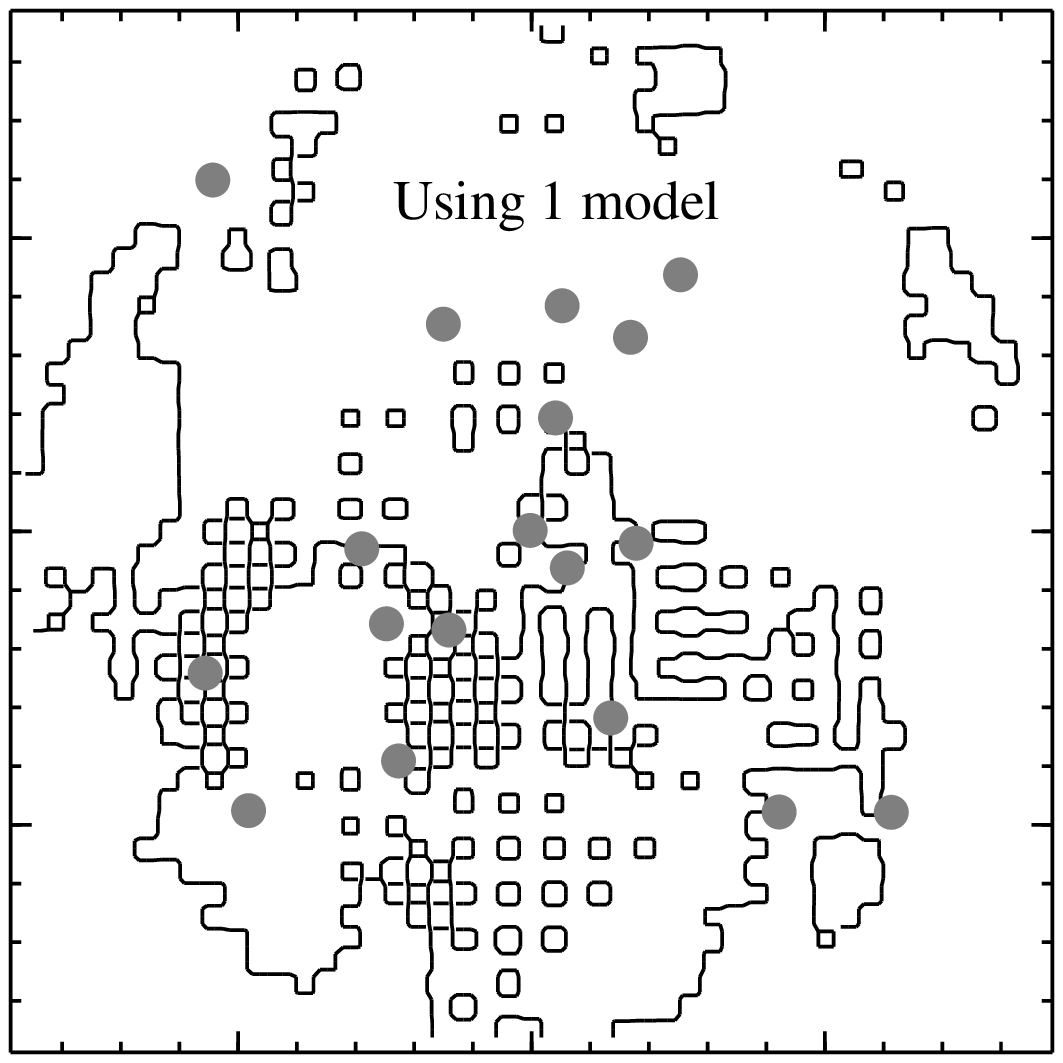}
\caption{Substructure in \snclus\ derived from subsets of the model
ensemble, for comparison with the full-ensemble result (i.e., using
400 models) shown in the upper panel in Fig.~\ref{J1004-rg}.  As in
the earlier figure the contours show an overdensity level of
$10^8M_\odot$.  The excess mass enclosed within the contours is 7--8\%
of the total mass, the latter being consistently within a few percent
of $10^{14}M_\odot$.}
\label{J1004-sub}
\end{figure}

The meso-structure results in the main part of this paper refer to the
average models from ensembles of 400.  It is interesting to also
examine smaller averages over the ensemble, down to individual models.
In Figure~\ref{J1004-sub} we show the inferred meso-structure in
\snclus\ if we take a single model, or average 4, 16, or 80 models,
and then remove the best-fit projected NFW profile in each case.  The
no-average case shows a lot of pixel-to-pixel variation, which
gradually fades as we take larger and larger averages.  The unaveraged
model still fits the data precisely, so the pixel-to-pixel variation
is not noise in the usual sense.  The variation appears and fades
because in any one model some pixels will be at the tails of their
respective probability distributions, whereas in an average all pixels
are at their expectation values.  Nevertheless, the meso-structure is
discernable even in a single model, and becomes quite clear even in an
average over 4 models.

We conclude that the meso-structure cannot be an artifact of
ensemble-averaging.

\newpage

\bibliographystyle{apj}
\bibliography{ms.bbl}

\begin{thebibliography}{47}
\expandafter\ifx\csname natexlab\endcsname\relax\def\natexlab#1{#1}\fi

\bibitem[{{Abdelsalam} {et~al.}(1998{\natexlab{a}}){Abdelsalam}, {Saha}, \&
  {Williams}}]{1998MNRAS.294..734A}
{Abdelsalam}, H.~M., {Saha}, P., \& {Williams}, L.~L.~R. 1998{\natexlab{a}},
  \mnras, 294, 734

\bibitem[{{Abdelsalam} {et~al.}(1998{\natexlab{b}}){Abdelsalam}, {Saha}, \&
  {Williams}}]{1998AJ....116.1541A}
---. 1998{\natexlab{b}}, \aj, 116, 1541

\bibitem[{{Adami} {et~al.}(2005){Adami}, {Biviano}, {Durret}, \&
  {Mazure}}]{2005A&A...443...17A}
{Adami}, C., {Biviano}, A., {Durret}, F., \& {Mazure}, A. 2005, \aap, 443, 17

\bibitem[{{Amara} {et~al.}(2006){Amara}, {Metcalf}, {Cox}, \&
  {Ostriker}}]{2006MNRAS.367.1367A}
{Amara}, A., {Metcalf}, R.~B., {Cox}, T.~J., \& {Ostriker}, J.~P. 2006, \mnras,
  367, 1367

\bibitem[{{Andersson} \& {Madejski}(2004)}]{2004ApJ...607..190A}
{Andersson}, K.~E. \& {Madejski}, G.~M. 2004, \apj, 607, 190

\bibitem[{{Bartelmann}(1996)}]{1996A&A...313..697B}
{Bartelmann}, M. 1996, \aap, 313, 697

\bibitem[{{Bartelmann} {et~al.}(1995){Bartelmann}, {Steinmetz}, \&
  {Weiss}}]{1995A&A...297....1B}
{Bartelmann}, M., {Steinmetz}, M., \& {Weiss}, A. 1995, \aap, 297, 1

\bibitem[{{Belokurov} {et~al.}(2006){Belokurov}, {Zucker}, {Evans}, {Gilmore},
  {Vidrih}, {Bramich}, {Newberg}, {Wyse}, {Irwin}, {Fellhauer}, {Hewett},
  {Walton}, {Wilkinson}, {Cole}, {Yanny}, {Rockosi}, {Beers}, {Bell},
  {Brinkmann}, {Ivezi{\'c}}, \& {Lupton}}]{2006ApJ...642L.137B}
{Belokurov}, V., {Zucker}, D.~B., {Evans}, N.~W., {Gilmore}, G., {Vidrih}, S.,
  {Bramich}, D.~M., {Newberg}, H.~J., {Wyse}, R.~F.~G., {Irwin}, M.~J.,
  {Fellhauer}, M., {Hewett}, P.~C., {Walton}, N.~A., {Wilkinson}, M.~I.,
  {Cole}, N., {Yanny}, B., {Rockosi}, C.~M., {Beers}, T.~C., {Bell}, E.~F.,
  {Brinkmann}, J., {Ivezi{\'c}}, {\v Z}., \& {Lupton}, R. 2006, \apjl, 642,
  L137

\bibitem[{{Brada{\v c}} {et~al.}(2004){Brada{\v c}}, {Schneider}, {Lombardi},
  {Steinmetz}, {Koopmans}, \& {Navarro}}]{2004A&A...423..797B}
{Brada{\v c}}, M., {Schneider}, P., {Lombardi}, M., {Steinmetz}, M.,
  {Koopmans}, L.~V.~E., \& {Navarro}, J.~F. 2004, \aap, 423, 797

\bibitem[{{Broadhurst} {et~al.}(2005){Broadhurst}, {Ben{\'{\i}}tez}, {Coe},
  {Sharon}, {Zekser}, {White}, {Ford}, {Bouwens}, {Blakeslee}, {Clampin},
  {Cross}, {Franx}, {Frye}, {Hartig}, {Illingworth}, {Infante}, {Menanteau},
  {Meurer}, {Postman}, {Ardila}, {Bartko}, {Brown}, {Burrows}, {Cheng},
  {Feldman}, {Golimowski}, {Goto}, {Gronwall}, {Herranz}, {Holden}, {Homeier},
  {Krist}, {Lesser}, {Martel}, {Miley}, {Rosati}, {Sirianni}, {Sparks},
  {Steindling}, {Tran}, {Tsvetanov}, \& {Zheng}}]{2005ApJ...621...53B}
{Broadhurst}, T., {Ben{\'{\i}}tez}, N., {Coe}, D., {Sharon}, K., {Zekser}, K.,
  {White}, R., {Ford}, H., {Bouwens}, R., {Blakeslee}, J., {Clampin}, M.,
  {Cross}, N., {Franx}, M., {Frye}, B., {Hartig}, G., {Illingworth}, G.,
  {Infante}, L., {Menanteau}, F., {Meurer}, G., {Postman}, M., {Ardila}, D.~R.,
  {Bartko}, F., {Brown}, R.~A., {Burrows}, C.~J., {Cheng}, E.~S., {Feldman},
  P.~D., {Golimowski}, D.~A., {Goto}, T., {Gronwall}, C., {Herranz}, D.,
  {Holden}, B., {Homeier}, N., {Krist}, J.~E., {Lesser}, M.~P., {Martel},
  A.~R., {Miley}, G.~K., {Rosati}, P., {Sirianni}, M., {Sparks}, W.~B.,
  {Steindling}, S., {Tran}, H.~D., {Tsvetanov}, Z.~I., \& {Zheng}, W. 2005,
  \apj, 621, 53

\bibitem[{{Congdon} \& {Keeton}(2005)}]{2005MNRAS.364.1459C}
{Congdon}, A.~B. \& {Keeton}, C.~R. 2005, \mnras, 364, 1459

\bibitem[{{Dalal} \& {Kochanek}(2002)}]{2002ApJ...572...25D}
{Dalal}, N. \& {Kochanek}, C.~S. 2002, \apj, 572, 25

\bibitem[{{Diego} {et~al.}(2005){Diego}, {Sandvik}, {Protopapas}, {Tegmark},
  {Ben{\'{\i}}tez}, \& {Broadhurst}}]{2005MNRAS.362.1247D}
{Diego}, J.~M., {Sandvik}, H.~B., {Protopapas}, P., {Tegmark}, M.,
  {Ben{\'{\i}}tez}, N., \& {Broadhurst}, T. 2005, \mnras, 362, 1247

\bibitem[{{Duc} {et~al.}(2002){Duc}, {Poggianti}, {Fadda}, {Elbaz}, {Flores},
  {Chanial}, {Franceschini}, {Moorwood}, \& {Cesarsky}}]{2002A&A...382...60D}
{Duc}, P.-A., {Poggianti}, B.~M., {Fadda}, D., {Elbaz}, D., {Flores}, H.,
  {Chanial}, P., {Franceschini}, A., {Moorwood}, A., \& {Cesarsky}, C. 2002,
  \aap, 382, 60

\bibitem[{{Evans} \& {Witt}(2003)}]{2003MNRAS.345.1351E}
{Evans}, N.~W. \& {Witt}, H.~J. 2003, \mnras, 345, 1351

\bibitem[{{Ferreras} {et~al.}(2005){Ferreras}, {Saha}, \&
  {Williams}}]{2005ApJ...623L...5F}
{Ferreras}, I., {Saha}, P., \& {Williams}, L.~L.~R. 2005, \apjl, 623, L5

\bibitem[{{Fohlmeister} {et~al.}(2006){Fohlmeister}, {Kochanek}, \&
  {Falco}}]{fohlmeister}
{Fohlmeister}, J., {Kochanek}, C.~S., \& {Falco}, E. 2006, {\tt
  astro-ph/0607513}

\bibitem[{{Garrett} {et~al.}(1994){Garrett}, {Calder}, {Porcas}, {King},
  {Walsh}, \& {Wilkinson}}]{1994MNRAS.270..457G}
{Garrett}, M.~A., {Calder}, R.~J., {Porcas}, R.~W., {King}, L.~J., {Walsh}, D.,
  \& {Wilkinson}, P.~N. 1994, \mnras, 270, 457

\bibitem[{{Gorenstein} {et~al.}(1984){Gorenstein}, {Shapiro}, {Rogers},
  {Cohen}, {Corey}, {Porcas}, {Falco}, {Bonometti}, {Preston}, {Rius}, \&
  {Whitney}}]{1984ApJ...287..538G}
{Gorenstein}, M.~V., {Shapiro}, I.~I., {Rogers}, A.~E.~E., {Cohen}, N.~L.,
  {Corey}, B.~E., {Porcas}, R.~W., {Falco}, E.~E., {Bonometti}, R.~J.,
  {Preston}, R.~A., {Rius}, A., \& {Whitney}, A.~R. 1984, \apj, 287, 538

\bibitem[{{Gray} {et~al.}(2002){Gray}, {Taylor}, {Meisenheimer}, {Dye}, {Wolf},
  \& {Thommes}}]{2002ApJ...568..141G}
{Gray}, M.~E., {Taylor}, A.~N., {Meisenheimer}, K., {Dye}, S., {Wolf}, C., \&
  {Thommes}, E. 2002, \apj, 568, 141

\bibitem[{{Grogin} \& {Narayan}(1996)}]{1996ApJ...464...92G}
{Grogin}, N.~A. \& {Narayan}, R. 1996, \apj, 464, 92

\bibitem[{{Halkola} {et~al.}(2006){Halkola}, {Seitz}, \&
  {Pannella}}]{2006MNRAS.372.1425H}
{Halkola}, A., {Seitz}, S., \& {Pannella}, M. 2006, \mnras, 372, 1425

\bibitem[{{Inada} {et~al.}(2005){Inada}, {Oguri}, {Keeton}, {Eisenstein},
  {Castander}, {Chiu}, {Hall}, {Hennawi}, {Johnston}, {Pindor}, {Richards},
  {Rix}, {Schneider}, \& {Zheng}}]{2005PASJ...57L...7I}
{Inada}, N., {Oguri}, M., {Keeton}, C.~R., {Eisenstein}, D.~J., {Castander},
  F.~J., {Chiu}, K., {Hall}, P.~B., {Hennawi}, J.~F., {Johnston}, D.~E.,
  {Pindor}, B., {Richards}, G.~T., {Rix}, H.-W.~R., {Schneider}, D.~P., \&
  {Zheng}, W. 2005, \pasj, 57, L7

\bibitem[{{Inada} {et~al.}(2003){Inada}, {Oguri}, {Pindor}, {Hennawi}, {Chiu},
  {Zheng}, {Ichikawa}, {Gregg}, {Becker}, {Suto}, {Strauss}, {Turner},
  {Keeton}, {Annis}, {Castander}, {Eisenstein}, {Frieman}, {Fukugita}, {Gunn},
  {Johnston}, {Kent}, {Nichol}, {Richards}, {Rix}, {Sheldon}, {Bahcall},
  {Brinkmann}, {Ivezi{\'c}}, {Lamb}, {McKay}, {Schneider}, \&
  {York}}]{2003Natur.426..810I}
{Inada}, N., {Oguri}, M., {Pindor}, B., {Hennawi}, J.~F., {Chiu}, K., {Zheng},
  W., {Ichikawa}, S.-I., {Gregg}, M.~D., {Becker}, R.~H., {Suto}, Y.,
  {Strauss}, M.~A., {Turner}, E.~L., {Keeton}, C.~R., {Annis}, J., {Castander},
  F.~J., {Eisenstein}, D.~J., {Frieman}, J.~A., {Fukugita}, M., {Gunn}, J.~E.,
  {Johnston}, D.~E., {Kent}, S.~M., {Nichol}, R.~C., {Richards}, G.~T., {Rix},
  H.-W., {Sheldon}, E.~S., {Bahcall}, N.~A., {Brinkmann}, J., {Ivezi{\'c}}, {\v
  Z}., {Lamb}, D.~Q., {McKay}, T.~A., {Schneider}, D.~P., \& {York}, D.~G.
  2003, \nat, 426, 810

\bibitem[{{Keeton} \& {Winn}(2003)}]{2003ApJ...590...39K}
{Keeton}, C.~R. \& {Winn}, J.~N. 2003, \apj, 590, 39

\bibitem[{{Limousin} {et~al.}(2006){Limousin}, {Richard}, {Kneib}, {Jullo},
  {Fort}, {Soucail}, {El{\'{\i}}asd{\'o}ttir}, {Natarajan}, {Smail}, {Ellis},
  {Czoske}, {Hudelot}, {Bardeau}, {Ebeling}, \& {Smith}}]{2006astro.ph.12165L}
{Limousin}, M., {Richard}, J., {Kneib}, J.~., {Jullo}, E., {Fort}, B.,
  {Soucail}, G., {El{\'{\i}}asd{\'o}ttir}, A., {Natarajan}, P., {Smail}, I.,
  {Ellis}, R.~S., {Czoske}, O., {Hudelot}, P., {Bardeau}, S., {Ebeling}, H., \&
  {Smith}, G.~P. 2006, ArXiv Astrophysics e-prints

\bibitem[{{Macci{\`o}} \& {Miranda}(2006)}]{2006MNRAS.368..599M}
{Macci{\`o}}, A.~V. \& {Miranda}, M. 2006, \mnras, 368, 599

\bibitem[{{Mao} \& {Schneider}(1998)}]{1998MNRAS.295..587M}
{Mao}, S. \& {Schneider}, P. 1998, \mnras, 295, 587

\bibitem[{{Metcalf} \& {Zhao}(2002)}]{2002ApJ...567L...5M}
{Metcalf}, R.~B. \& {Zhao}, H. 2002, \apjl, 567, L5

\bibitem[{{Oguri} {et~al.}(2004){Oguri}, {Inada}, {Keeton}, {Pindor},
  {Hennawi}, {Gregg}, {Becker}, {Chiu}, {Zheng}, {Ichikawa}, {Suto}, {Turner},
  {Annis}, {Bahcall}, {Brinkmann}, {Castander}, {Eisenstein}, {Frieman},
  {Goto}, {Gunn}, {Johnston}, {Kent}, {Nichol}, {Richards}, {Rix}, {Schneider},
  {Sheldon}, \& {Szalay}}]{2004ApJ...605...78O}
{Oguri}, M., {Inada}, N., {Keeton}, C.~R., {Pindor}, B., {Hennawi}, J.~F.,
  {Gregg}, M.~D., {Becker}, R.~H., {Chiu}, K., {Zheng}, W., {Ichikawa}, S.-I.,
  {Suto}, Y., {Turner}, E.~L., {Annis}, J., {Bahcall}, N.~A., {Brinkmann}, J.,
  {Castander}, F.~J., {Eisenstein}, D.~J., {Frieman}, J.~A., {Goto}, T.,
  {Gunn}, J.~E., {Johnston}, D.~E., {Kent}, S.~M., {Nichol}, R.~C., {Richards},
  G.~T., {Rix}, H.-W., {Schneider}, D.~P., {Sheldon}, E.~S., \& {Szalay}, A.~S.
  2004, \apj, 605, 78

\bibitem[{{Raychaudhury} {et~al.}(2003){Raychaudhury}, {Saha}, \&
  {Williams}}]{2003AJ....126...29R}
{Raychaudhury}, S., {Saha}, P., \& {Williams}, L.~L.~R. 2003, \aj, 126, 29

\bibitem[{{Ros} {et~al.}(2000){Ros}, {Guirado}, {Marcaide}, {P{\'e}rez-Torres},
  {Falco}, {Mu{\~n}oz}, {Alberdi}, \& {Lara}}]{2000A&A...362..845R}
{Ros}, E., {Guirado}, J.~C., {Marcaide}, J.~M., {P{\'e}rez-Torres}, M.~A.,
  {Falco}, E.~E., {Mu{\~n}oz}, J.~A., {Alberdi}, A., \& {Lara}, L. 2000, \aap,
  362, 845

\bibitem[{{Saha}(2003)}]{2003pda.book}
{Saha}, P. 2003, {Principles of Data Analysis} (Great Malvern, UK, Cappella
  Archive, 2003)

\bibitem[{{Saha} {et~al.}(2006{\natexlab{a}}){Saha}, {Coles}, {Macci{\`o}}, \&
  {Williams}}]{2006ApJ...650L..17S}
{Saha}, P., {Coles}, J., {Macci{\`o}}, A.~V., \& {Williams}, L.~L.~R.
  2006{\natexlab{a}}, \apjl, 650, L17

\bibitem[{{Saha} {et~al.}(2006{\natexlab{b}}){Saha}, {Read}, \&
  {Williams}}]{2006ApJ...652L...5S}
{Saha}, P., {Read}, J.~I., \& {Williams}, L.~L.~R. 2006{\natexlab{b}}, \apjl,
  652, L5

\bibitem[{{Saha} \& {Williams}(1997)}]{1997MNRAS.292..148S}
{Saha}, P. \& {Williams}, L.~L.~R. 1997, \mnras, 292, 148

\bibitem[{{Saha} \& {Williams}(2003)}]{2003AJ....125.2769S}
---. 2003, \aj, 125, 2769

\bibitem[{{Saha} \& {Williams}(2004)}]{2004AJ....127.2604S}
---. 2004, \aj, 127, 2604

\bibitem[{{Saha} \& {Williams}(2006)}]{2006ApJ...653..936S}
---. 2006, \apj, 653, 936

\bibitem[{{Saha} {et~al.}(2001){Saha}, {Williams}, \&
  {Abdelsalam}}]{2001ASPC..237..279S}
{Saha}, P., {Williams}, L.~L.~R., \& {Abdelsalam}, H.~M. 2001, in ASP Conf.
  Ser. 237: Gravitational Lensing: Recent Progress and Future Go, ed. T.~G.
  {Brainerd} \& C.~S. {Kochanek}, 279--+

\bibitem[{{Sharon} {et~al.}(2005){Sharon}, {Ofek}, {Smith}, {Broadhurst},
  {Maoz}, {Kochanek}, {Oguri}, {Suto}, {Inada}, \&
  {Falco}}]{2005ApJ...629L..73S}
{Sharon}, K., {Ofek}, E.~O., {Smith}, G.~P., {Broadhurst}, T., {Maoz}, D.,
  {Kochanek}, C.~S., {Oguri}, M., {Suto}, Y., {Inada}, N., \& {Falco}, E.~E.
  2005, \apjl, 629, L73

\bibitem[{{Trotter} {et~al.}(2000){Trotter}, {Winn}, \&
  {Hewitt}}]{2000ApJ...535..671T}
{Trotter}, C.~S., {Winn}, J.~N., \& {Hewitt}, J.~N. 2000, \apj, 535, 671

\bibitem[{{Williams} \& {Saha}(2004)}]{2004AJ....128.2631W}
{Williams}, L.~L.~R. \& {Saha}, P. 2004, \aj, 128, 2631

\bibitem[{{Witt} {et~al.}(2000){Witt}, {Mao}, \&
  {Keeton}}]{2000ApJ...544...98W}
{Witt}, H.~J., {Mao}, S., \& {Keeton}, C.~R. 2000, \apj, 544, 98

\bibitem[{{Yoo} {et~al.}(2005){Yoo}, {Kochanek}, {Falco}, \&
  {McLeod}}]{2005ApJ...626...51Y}
{Yoo}, J., {Kochanek}, C.~S., {Falco}, E.~E., \& {McLeod}, B.~A. 2005, \apj,
  626, 51

\bibitem[{{Yoo} {et~al.}(2006){Yoo}, {Kochanek}, {Falco}, \&
  {McLeod}}]{2006ApJ...642...22Y}
---. 2006, \apj, 642, 22

\bibitem[{{Zekser} {et~al.}(2006){Zekser}, {White}, {Broadhurst},
  {Ben{\'{\i}}tez}, {Ford}, {Illingworth}, {Blakeslee}, {Postman}, {Jee}, \&
  {Coe}}]{2006ApJ...640..639Z}
{Zekser}, K.~C., {White}, R.~L., {Broadhurst}, T.~J., {Ben{\'{\i}}tez}, N.,
  {Ford}, H.~C., {Illingworth}, G.~D., {Blakeslee}, J.~P., {Postman}, M.,
  {Jee}, M.~J., \& {Coe}, D.~A. 2006, \apj, 640, 639

\end{thebibliography}

\end{document}